\documentclass[11pt,a4paper]{article}
 
\usepackage{textcomp}
\usepackage{pdfsync}

\usepackage{amssymb}
\usepackage{amsmath}

\usepackage{graphicx}
\usepackage{latexsym}
\usepackage{appendix}
\usepackage{srcltx}
\def\CD{{\cal D}}
\def\CE{{\cal E}}

\def\CK{{\cal K}}
\def\CN{{\cal N}}
\def\CO{{\cal O}}

\def\CS{{\cal S}}

\def\CW{{\cal W}}

\def\centeron#1#2{{\setbox0=\hbox{#1}\setbox1=\hbox{#2}\ifdim
   \wd1>\wd0\kern.48\wd1\kern-.48\wd0\fi
   \copy0\kern-.48\wd0\kern-.48\wd1\copy1\ifdim\wd0>\wd1
   \kern.48\wd0\kern-.48\wd1\fi}}

\def\JHEP{JHEP~}

\def\PRL{Phys. Rev. Lett.~}
\def\PR {Phys. Rev.~}
\def\CQG {Class. Quant. Grav.~}

\newcommand{\beq}{\begin{equation}}
\newcommand{\eeq}{\end{equation}}
\newcommand{\bea}{\begin{eqnarray}}
\newcommand{\eea}{\end{eqnarray}}
\newcommand{\ba}{\begin{array}}
\newcommand{\ea}{\end{array}}

\newcommand{\p}{\partial}
\newcommand{\nn}{\nonumber}

\newcommand{\half}{\frac{1}{2}}

\renewcommand{\thefootnote}{\arabic{footnote}}

\begin{document}

\hskip11.6cm{CQUeST--2012-0549}
\vskip1cm

\begin{center}
 \LARGE \bf
 Extremal Black Holes and Holographic C-Theorem
\end{center}

\vskip1.5cm

\centerline{\large
  Yongjoon Kwon$^1$, ~
 Soonkeon Nam$^2$, ~
 Jong-Dae Park$^3$, ~  Sang-Heon
Yi$^4$ }

\hskip1cm

\begin{quote}{
Department of Physics and Research Institute of
Basic Science, Kyung Hee University,
Seoul 130-701, Korea$^{1,2,3}$ \\
Center for Quantum Spacetime, Sogang University, Seoul 121-741,
Korea$^4$}\\
\end{quote}
\hskip2cm

\centerline{\large \bf Abstract}
\hskip0.3cm

We found Bogomol'nyi type of the first order differential equations
in three dimensional Einstein gravity and the effective second order
ones in new massive gravity when an  interacting scalar field is
minimally coupled. Using these equations in Einstein gravity, we
obtain analytic solutions corresponding to extremally rotating
hairy black holes. We also obtain perturbatively  extremal black hole
solutions in new massive gravity using these lower order differential equations. All these solutions have the
anti de-Sitter spaces as their asymptotic geometries and as the
near horizon ones. This feature of solutions interpolating two
anti de-Sitter spaces leads to the construction of holographic
c-theorem in these cases. Since our lower order equations reduce
naturally to the well-known equations for domain walls, our
results can be regarded as the natural extension of domain walls
to more generic cases.

\hskip2cm

\noindent\underline{\hskip12cm}\\
 {\it \small
$^1$emwave@khu.ac.kr, $^2$nam@khu.ac.kr, $^3$jdpark@khu.ac.kr,
$^4$shyi@sogang.ac.kr}

\thispagestyle{empty}
\renewcommand{\thefootnote}{\arabic{footnote}}
\setcounter{footnote}{0}

\newpage


\section{Introduction}

Recently, there has been much interest in $c$-theorem in various dimensions. The content of  the theorem is that a certain central charge
function, which is regarded as counting the number of degrees of freedom, should be a monotonically
decreasing function along a Wilsonian renormalization group(RG) flow. The interesting recent developments include the formal `proof'~\cite{Komargodski:2011vj} of
four-dimensional $c$-(or $a$-)theorem long after its
conjecture~\cite{Cardy:1988cwa}, the discovery of the relation
between central charge and the entanglement
entropy(EE)~\cite{Casini:2011kv}, and the identification of the
conjectured free energy maximization in three-dimensional field theory named as F-maximization with
the internal
space volume
minimization known as Z-minimization~\cite{Jafferis:2010un}.
Furthermore, the content of this theorem was constructed
holographically in Einstein gravity through the AdS/CFT
correspondence~\cite{Freedman:1999gp,de Boer:1999xf} and
extended recently to higher derivative
gravity~\cite{Myers:2010xs}.
The pioneering work on all these developments in c-theorem was
done
by Zamolodchikov a few decades ago~\cite{Zamolodchikov:1986gt},
who
argued that c-theorem is natural intuitively in the sense of RG
flow
from UV  to IR and proved rigorously that it holds  in two
dimensions under the assumption on generic properties of field theory like unitarity, conformal
invariance, etc.

In general, it seems natural that a kind of $c$-theorem holds in
any
kind of sensible unitary field theories. Contrary to this simple intuition, it is challenging
to prove this
theorem
because its nature requires essentially the non-perturbative
method
and furthermore its proof or verification depends strongly on
the
spacetime dimensions, in the works that have been done so far. There have been some attempts to overcome this situation. For instance,
the
relation between central charge and EE was shown in two
dimensions,
and then two-dimensional c-theorem is rederived by using the
property of EE. This construction is extended to higher
dimensions
and argued to be a generic proof of c-theorem. Other
attempts
are the holographic construction of $c$-theorem in various
gravity
models, which reveal that the conjectured $c$-theorem is
consistent
with the holographic construction. In four dimensions there are
two
central charges called as $a$ and $c$, which are the
coefficients of
 Euler and  Weyl  density in the trace anomaly formula. It was
conjectured and proved recently that $a$ is the relevant central charge consistent with the monotonic flow
property. The
holographic
construction is very appealing since it reproduces this result
nicely and it can be extended to various other dimensions
including
odd ones.

Though the holographic construction is very useful to understand c-theorem uniformly in various
dimensions, its
construction is
restricted usually to the simplest form of domain wall metric in
the gravity side. 
In the
context of the AdS/CFT correspondence, domain wall solutions in
gravity with two asymptotic AdS spaces correspond to RG flow
trajectory between two conformal points in dual field theory. In this holographic construction,
appropriate central
charge flow functions, which coincide with central charges at the two
conformal
points, are constructed by using metric functions. And then
their
monotonicity is verified through the equations of motion and
null
energy condition on matters which is imposed as a sensible
condition
in the gravity side.  This is the content of the so-called
holographic $c$-theorem.

Even though this holographic $c$-theorem may be checked without
the
explicit domain wall solutions, it is more satisfactory to
obtain
the analytic domain wall solutions consistent with   holographic $c$-theorem. Interestingly, it has
been shown that
the domain
wall
metric in Einstein gravity satisfies Bogomol'nyi type of the
first
order differential equations which were derived by minimizing a certain energy functional through
complete
squares~\cite{Skenderis:1999mm} or by introducing a certain
`fake'
supersymmetry or supergravity~\cite{DeWolfe:1999cp,Low:2000pq}. Using these
first order equations, various analytic forms of domain wall
solutions have been obtained and shown to be consistent with  
holographic $c$-theorem  in various dimensions.

One may ask whether this domain wall geometry is the unique
candidate as the dual to the RG flow in boundary field theory.
It is rather clear that any geometry with two asymptotic AdS
spaces is viable as the dual to the RG flow and as the
background for  holographic $c$-theorem. However, it is not
easy to obtain such non-trivial background geometry analytically
in gravity since matters play important roles and hinder
analytic treatments. In this regard, three-dimensional gravity
is exceptional since various analytic solutions are found
including black holes with scalar hairs~\cite{Henneaux:2002wm}.
Indeed, there is the realization of this idea by using more
complicated three-dimensional geometry than domain walls, which
turns out to be consistent with   holographic
$c$-theorem~\cite{Hotta:2008xt}. The relevant geometry is given
by the extremally rotating three-dimensional AdS black holes
which interpolate between AdS space at the asymptotic infinity
and near horizon AdS geometry.

The existence of the analytic black hole solutions allows  the
explicit realization of holographic RG flow via Hamilton-Jacobi
formalism and the check of  the  holographic $c$-theorem. However,
the shortcoming in these extremal black hole solutions given
in Ref.~\cite{Hotta:2008xt}, compared to the domain wall solutions,
is
the fact that one needs to make a specific `ad hoc' choice of
the
scalar potential to obtain analytic results. This point is even
amplified when one considers higher curvature gravity like new
massive gravity(NMG)~\cite{Bergshoeff:2009hq}  which is recently
introduced as a non-linear completion
of
Pauli-Fierz linear massive graviton theory and shown to be
consistent with a simple form of a  holographic
$c$-theorem~\cite{Sinha:2010ai}. It becomes very difficult to
choose
`ad hoc' scalar potential in the NMG case, which is contrasted
to
the fact  that domain walls satisfy first order differential
equations even in NMG and allow  analytic
results~\cite{Camara:2010cd}.

In the context of the AdS/CFT correspondence, the generic nature
of
holographic construction seems to imply that there exists a more unified and systematic approach to
these extremal
black holes
with
two AdS asymptotics. It is natural to suspect the existence of
some reduced differential equations for extremal black holes as
for
domain walls. One of main results in this paper is the discovery
of
such differential equations for three-dimensional AdS black
holes
in Einstein gravity and in NMG. We also show that such equations
are enough for the consistency with the holographic $c$-theorem
when a
certain central charge flow function is chosen.

This paper is organized as follows. In the next section we find
the
Bogomol'nyi type of first order differential equations,  which
solves the full equations of motion, in three-dimensional
Einstein
gravity interacting with a scalar field. It turns out that these
restricted first order equations of motion represent extremally rotating black holes with scalar
hairs. By solving
these first
order equations of motion in a more or less systematic way, we
obtain some analytic hairy black hole solutions which include
the
case given in Ref.~\cite{Hotta:2008xt} as a special case. In section
three we consider new massive gravity as another gravity theory
to
obtain reduced differential equations for extremal AdS black
holes.
As in the Einstein gravity case, it is shown that Bogomol'nyi
type
of lower order differential equations can be obtained which
include
domain wall solutions as special cases. By solving these
equations
asymptotically, we show that there are extremally rotating black hole solutions consistent with a
holographic
$c$-theorem. In
section
four, we consider the holographic $c$-theorem in our setup and show that it holds generically by using
reduced lower
order equations
of
motions. In the final section, we summarize our results with
some
comments and discuss open issues.

\section{Extremal Black Hole Solutions in Einstein Gravity}
In this section we consider three-dimensional Einstein gravity
with a
minimally coupled interacting scalar field. We find Bogomol'nyi type of first order differential
equations which
solve full
equations of motion.  This  can be regarded as the extension of first order equations for domain
walls~\cite{Skenderis:1999mm,DeWolfe:1999cp} to more
generic
cases.   It turns out that the simplest solutions of these
equations, which are given by a constant scalar field, correspond
to
extremal BTZ black holes~\cite{Banados:1992wn}. After showing
that
these equations describe the extremally rotating black holes, we obtain analytic solutions of some hairy black holes in a systematic way.

\subsection{First order equations of motion}
In the convention of mostly plus signs for the metric with  the convention of 
curvature tensors  as $[\nabla_{\mu}\, \nabla_{\nu}
]V_{\rho}
= R_{\mu\nu\rho\sigma}V^{\sigma}$ and  $R_{\mu\nu} =
g^{\alpha\beta}R_{\alpha\mu\beta\nu}$, our starting action for
Einstein gravity with a minimally coupled scalar field is given
by
\begin{equation}\label{NMG}
S =\frac{1}{16\pi G}\int d^3x\sqrt{-g}\bigg[ R -\half
\p_{\mu}\phi \p^{\mu}\phi -V(\phi) \bigg]\,,
\end{equation}
of which the equations
of motions(EOM)  are composed of scalar field equation and the metric field equations  as follows ;
\beq 0 = E_{\phi} \equiv  \nabla^2\phi - \frac{\p V}{\p \phi}\,, \qquad 
0=E_{\mu\nu} \equiv \CE_{\mu\nu} - T_{\mu\nu} \,,    
 \eeq
where
\[ \CE_{\mu\nu} \equiv R_{\mu\nu} -\half R g_{\mu\nu} \,, \qquad T_{\mu\nu} \equiv   \half \p_{\mu}\phi \p_{\nu}
\phi  -\half g_{\mu\nu} \Big[ \half\p_{\alpha}\phi\p^{\alpha}\phi +   V(\phi)\Big]\,.  \]\

To find asymptotically AdS black hole solutions with an
interacting scalar field in three dimensions, let us take our
metric ansatz in AdS-Schwarzschild-like coordinates as
\begin{eqnarray}\label{BTZ}
ds^2 = L^2 \Big[ -e^{2A(r)} dt^2 + e^{2B(r)}dr^2
+r^2\Big(d\theta + e^{C(r)}dt\Big)^2
 \Big] \,,
\end{eqnarray}
where $L$ denotes the radius of asymptotic AdS space.
Asymptotically AdS black holes in
these coordinates mean that the asymptotic conditions on the functions $A(r),B(r),C(r)$
are given as follows;
\beq e^{A(r)}\Big|_{r\rightarrow\infty} \rightarrow r\,, \qquad
e^{B(r)}\Big|_{r\rightarrow\infty} \rightarrow \frac{1}{r}\,,
\qquad e^{C(r)}\Big|_{r\rightarrow\infty } \rightarrow const. +
\CO\Big(\frac{1}{r^2}\Big)\,. \eeq
Note that these boundary conditions are Brown-Henneaux type which
allow us to apply the standard central charge extraction by
Brown-Henneaux method \cite{Brown:1986nw}.

The equations of motion even in this case turn out to be
complicated non-linear differential equations. For instance, the
EOM for the scalar field is given by
\beq 0 =  E_{\phi} \equiv \frac{1}{L^2}\, e^{-2B}\Big[\Big(A'-B' +
\frac{1}{r}\Big) \phi' +\phi''\Big] - \frac{\p V}{\p
\phi}\,,\label{ScEOM} \eeq
where $'$ denotes  differentiation with respect to the radial
coordinate $r$.  The EOM for metric, $0=E_{\mu\nu}$, are relegated to appendix A.
To obtain analytic solutions of complicated
full EOM, it is very convenient to introduce the so-called
`superpotential' method which is originally applied to the domain
wall solutions. Historically, the terminology of superpotential
is chosen in analogy with supergravity expression for a scalar
potential.
When the scalar potential is represented by the so-called
superpotential $\CW$ as
\beq V(\phi) = \frac{1}{2L^2} \Big(\frac{\p \CW}{\p \phi}\Big)^2
- \frac{1}{2L^2} \CW^2\,, \eeq
the appropriate first order differential equations, which solve full
EOM,
are given by
\bea    \label{FirstDQ}
\phi'  &=&   -   e^{B} \frac{\p \CW}{\p \phi}\,,     \qquad
  A' =  e^{B}\CW -\frac{1}{r}\,, \\
A' + B' &=& \frac{r}{2}\, e^{2B}\Big(\frac{\p \CW}{\p
\phi}\Big)^2= - \frac{r}{2}\, e^{B}\Big(\frac{\p \CW}{\p
\phi}\Big)\, \phi' = \frac{r}{2}\phi'^2\,, \nn \\
(e^{C})' &=& \mp \frac{1}{r} e^{A}\Big( e^{B}\CW -
\frac{2}{r}\Big) = \mp \Big(\frac{1}{r}\, e^{A}\Big)' \,.\nn
\eea
These are motivated by similar expression in the domain wall
case.
This form of  differential equations, which we call as reduced
EOM,
is a considerable simplification compared to the original EOM,
though restricted solutions among all possible ones can be
obtained
from these reduced EOM.  Specifically, the last equation can be solved as
\beq e^{C} = C_{\mp}  \mp  \frac{1}{r}e^{A} \,,  \eeq
where the integration constants $C_{\mp}$ can take any values
consistently with asymptotic AdS space.

As a trivial solution of our first order equations, let us
consider the constant potential case, $V= - 2/L^2$ with a constant
scalar field. In this case superpotential is given by $\CW = 2$
and then one obtains
\beq  A' = 2e^{B}-\frac{1}{r}\,, \qquad A'+B'=0\,,  \eeq
which, with boundary conditions,  leads to the following solution;
\beq
e^{A}=e^{-B} = r - \frac{r^2_H}{r}\,, \qquad e^{C}= (C_{\mp} \mp
1) \pm \frac{r^2_H}{r^2}\,. \eeq
These metric functions represent  the  extremal BTZ
black holes. Although $C_{\mp}=\pm 1$ corresponds to the most
familiar form of extremal BTZ black holes, any value of the
constant, $C_{\mp}$, leads to the extremal BTZ black holes. In
fact, it is more useful to take $C_{\mp} = 0$ to simplify some
computations in our case\footnote{However,  when we consider conserved
charges  of black holes, which depend on the coordinates, we return  to
the coordinates $C_{\mp} = \pm1$ to reproduce the
standard form of conserved charges.}.
To see the convenience of this choice, let us consider the near
horizon geometry of extremal BTZ black holes given by
$r\rightarrow r_{H}$. Using a new radial coordinate $ \rho =
4(r -r_{H})$, one can easily identify the metric
of the near horizon geometry as
\[ ds^2_{NH} = \frac{L^2}{4}\left[ -\rho^2 dt^2 +
\frac{1}{\rho^2}d\rho^2 + 4r^2_H \Big(d\theta \mp
\frac{\rho}{2r_{H}}dt\Big)^2\right]\,, \]
which is the well-known metric form~\cite{Coussaert:1994tu} of
the
self-dual orbifold of $AdS_3$ with the radius $L$. Note that
this
geometry leads to zero Hawking temperature and so dual field
theory
to extremal BTZ can be thought to be at zero temperature. This
near
horizon geometry is interpreted as dual to the discrete light cone
quantization(DLCQ) of two-dimensional conformal field
theory(CFT)
and is  related to chiral two-dimensional
CFT~\cite{Balasubramanian:2009bg}.

{}From now on we will choose the integration constant as
$C_{\mp} =0$ so that $C$ is given by
\beq e^C = \mp \frac{1}{r}  e^A\,.  \label{MetricC}\eeq
%
The usual choice of
$C_{\mp} =
\pm 1$ can be recovered by a simple coordinate transformation:
$\theta \rightarrow \theta + C_{\mp}t$.

One may expect that all the solutions of our first order
equations correspond to some kind of extremal black holes as can
be inferred by the fact that the trivial solutions represent the
extremal BTZ black holes. This expectation is also natural in
analogy with charged extremal black hole solutions in
supergravity, in which those black holes are described by first
order equations which can be derived by Killing spinor
equations. As in the case of domain walls, we anticipate that
some `fake' Killing spinor equations might lead to our first
order equations. In the next section we present perturbative
analysis of these first order equations and show that black hole
solutions are indeed extremal.

\subsection{Extremally rotating black holes}
It is very convenient in solving the first order reduced
EOM to take $\phi$ as coordinates and $r$ as a function of
$\phi$, instead of the original form. Then the first
order  reduced EOM can be rewritten as
\bea &&\p_{\phi}(A+B)\, \p_{\phi}(\ln r) = \half \,,
\label{IntFOE} \\
&&   e^{B} \, \p_{\phi}   r\,   \p_{\phi}\CW = - 1\,, \nn    \\
&& \p_{\phi} (A + \ln r)\, \p_{\phi} (\ln \CW) =-1\,, \nn
\eea
where $\p_{\phi}$ denotes  differentiation with respect to
$\phi$
variable. One can see that the first two (or last two) equations
can
be immediately integrated in terms of $r$ and $\CW$ and lead to solutions of metric functions $A$ and $B$ as
\bea A & =& \frac{1}{2}\int^{\phi} d\phi' \Big[ \p_{\phi'} \ln
r(\phi') \Big]^{-1} + \ln \Big[-\p_{\phi} r\,
\p_{\phi}\CW\Big] 
= -\ln r +\int^{\phi}d\phi'\, e^B\,\CW\, \p_{\phi'}r\,,
 \nn \\
B &=& - \ln \Big[-  \p_{\phi} r \,\p_{\phi}\CW\Big] \,.
     \nn \eea
By inserting the expression of $A$ and $B$ functions in the
remaining equation, one obtains the differential equation for
$r(\phi)$ or for $\CW(\phi)$ as
\beq 0 = \Big[ \p^2_{\phi}\CW + \CW\Big] \p_{\phi}\ln r + \Big[
2(\p_{\phi} \ln r)^2 + \p^2_{\phi}\ln r +
\frac{1}{2}\Big]\p_{\phi}\CW
= \frac{1}{2r^2}\,\p_{\phi}\Big[ r^2\CW + (\p_{\phi}r^2)
(\p_{\phi}\CW)\Big]\,, \label{REOMDiff} \eeq
which can be integrated as
\beq r^2\CW + (\p_{\phi}r^2) (\p_{\phi}\CW)=constant \equiv
\Delta_0\,. \label{IntREOMDiff} \eeq
The physical meaning of this constant $\Delta_0$ will be given
shortly after discussing the near horizon geometry of our black
hole solutions, which will also  be related to the conserved charges.
One can see that the metric functions $A$ and $B$ are now
completely determined, in terms of the superpotential $\CW$ and
the constant $\Delta_0$, as
\beq
A = \ln r -\Delta_0\int^{\phi} d\phi'~ e^{B(\phi')}\,
\p_{\phi'}\Big(\frac{1}{r} \Big) \,, \qquad e^{-B} =
\frac{r}{2}\Big[\CW - \frac{1}{r^2}\Delta_0 \Big]\,.
\label{MetricB}
\eeq
This form of the metric function $B$ shows us explicitly that
$e^{-B}$ outside the horizon is a regular function of $(r -r_H)$
as long as $\CW$ can be written as a regular function of
$(r-r_H)$.

Before going ahead to present the analytic form of some hairy
AdS black hole solutions, let us consider the asymptotic and
near horizon behaviors of black hole solutions. According to the
boundary conditions and physical consideration, one may take as
\bea A(r) &=& \ln r  + \frac{a_1}{r^2} +  \cdots\,, \\
   B(r) &=& -\ln r   + \frac{b_1}{r^2} + \cdots\,,    \nn \\
\ln \CW &=& \ln 2 + \omega_1\, r^{-n} +  \cdots\,, \nn \\
\phi(r) &=& \phi_{\infty} + \frac{\phi_1}{r^k}   + \cdots\,, \nn
       \eea
where $\phi_\infty$ denotes the value of scalar field $\phi$ at
the asymptotic infinity $r=\infty$.
By solving the above differential equation (\ref{IntREOMDiff})
with (\ref{MetricB}) perturbatively, one can see that $n=2$,
$k=1$ and obtain
\bea a_1 &=& - \frac{1}{2}\Delta_0 \,, \quad b_1 = -\omega_1 +
\frac{1}{2}\Delta_0\,, \\
\ln \CW &=& \ln 2 + \frac{1}{4} (\phi - \phi_\infty)^2 +
\cdots\,, \nn \\
r^{-2} &=& \frac{1}{4\omega_1}(\phi-\phi_\infty)^2 + \cdots \,,
\nn
\eea
where the last equation is obtained by inverting $r$ as a
function of $\phi$ and using $\phi^2_1 = 4\omega_1$.
Note  that
the superpotential value outside the horizon becomes always
greater than its asymptotic value: $\CW(\phi) \ge
\CW(\phi_{\infty}) (=2)$. This fact can be checked explicitly
from the analytic solutions presented in the following. Even in
these perturbative solutions,
the power of the first order reduced EOM shows up as the complete
determination of $a_2$ and $b_2$ in terms of $\omega_1$ and
$\Delta_0$. Contrary to this, only the combination of $a_1+b_1=
-\omega_1$ is determined perturbatively by the original second
order EOM.

Now, we would like to show that the first order reduced EOM
represent extremal black holes by analyzing near horizon
geometry of black hole solutions of the reduced EOM. By assuming
the existence of the horizon, which is given by $e^{-B(r_H)}
=0$, with the expression of the metric function $B$ in
Eq.~(\ref{MetricB}), one can see that the constant $\Delta_0$ in
Eq.~(\ref{IntREOMDiff}) is determined as
\beq     \Delta_0 = r^2_H\CW(\phi_H)\,.  \eeq
By solving  Eq.~(\ref{IntREOMDiff}) perturbatively near the
horizon, one can see that 
the superpotential $\CW$ and the radial coordinate can be taken
as regular functions of $\phi$ as
\bea \CW (\phi) &=& \CW(\phi_H) - \frac{1}{2}\CW(\phi_H)\, (\phi
- \phi_H)^2 + \cdots\,, \\
  r(\phi) &=&  r_H +  h_0~ (\phi-\phi_H) + \cdots \,, \nn \eea
where $h_0$ is a certain constant\footnote{Here, we have assumed
$h_0\neq 0$. If this is not the case, the perturbative expression
should be modified.}. Accordingly,
metric functions $A$ and $B$   are  given by 
Eq.~(\ref{MetricB}) as
\bea  e^{A(r)} &=&  s_0\CW(\phi_H)(r-r_H) +\cdots\,, \\
e^{B(r)}  &=&   \frac{1}{ \CW(\phi_H)(r -r_H)} + \cdots\,, \nn\eea
where $s_0$ is a certain non-vanishing constant  and $\cdots$ denotes some regular functions of $(r-r_H)$.
Note that the constant $s_0$ is related to the interval of integration  in the expression of $A$ given in Eq.~(\ref{MetricB}). This form of the metric function $B$ on the near horizon 
shows us the extremality of black hole solutions of our first
order reduced EOM:
\beq
e^{-2B}\Big|_{r=r_H} =0\,, \qquad \frac{d}{dr} e^{-2B}
\Big|_{r=r_H} =0\,. \label{ExtCond}
\eeq

By introducing a new radial coordinate $\rho \equiv s_0\CW^2(\phi_H) ( r-r_H)
$, which is appropriate on the near horizon
region,
one may identify  the near horizon geometry as
\beq ds^2_{NH} = \Big[\frac{L}{\CW(\phi_H)}\Big]^2\left[ -\rho^2
dt^2 + \frac{1}{\rho^2}d\rho^2 + \hat{r}^2_H\Big(d\theta \mp
\frac{\rho}{\hat{r}_H}dt\Big)^2\right], \quad \hat{r}_H \equiv
r_H \CW(\phi_H)\,.\label{NHM} \eeq
One may introduce the AdS scale $\bar{L}$ on the near horizon
geometry through $V(\phi_H) = -2/\bar{L}^2$, which is related to
the superpotential as $V(\phi_H) = -\CW^2(\phi_H)/2L^2$ since
$\p_{\phi}\CW(\phi_H)=0$ at the horizon. Interestingly, in terms
of this scale $\bar{L}$ and $\bar{r}_H=\hat{r}_H/2 $  the near horizon geometry may be written
in the form of
\beq ds^2_{NH} = \frac{\bar{L}^2}{4}\left[ -\rho^2 dt^2 +
\frac{1}{\rho^2}d\rho^2 + 4\bar{r}^2_H\Big(d\theta \mp
\frac{\rho}{2\bar{r}_H}dt\Big)^2\right]\,, \qquad \bar{r}_H
\equiv r_H \Big[ \frac{L }{ \bar{L}}\Big] \,, \label{NHM} \eeq
which is just the metric  for the
self-dual
orbifold of $AdS_3$ with the radius $\bar{L}$. Now, one can
identify
$\bar{L}$ with the superpotential value at the horizon
$\CW(\phi_H)$
or the constant $\Delta_0$ as
\beq {\bar L} = \frac{2L}{\CW(\phi_H)} =
\frac{2r^2_HL}{\Delta_0}\,. \label{barL}\eeq
This explains the physical meaning of the integration constant
$\Delta_0$, which is related to the information about the near
horizon geometry. This result will be consistent with  the 
holographic $c$-theorem, as will be discussed in the section 4.

Now, let us consider some physical quantities related to these
extremal black holes. One will see that  $\Delta_0$ is directly related to the conserved charges of black holes. Using
the above
explicit asymptotic
expressions of the metric and the scalar field, one can obtain
masses
and angular momenta of black holes, for instance, through the so-called
Abbott-Deser-Tekin(ADT) method~\cite{Abbott:1981ff}. Note that, in this ADT approach,
one
does not need to compute    contributions separately  from the
metric and the scalar field, which is contrary to the
quasi-local
charge method given
in Refs.~\cite{Henneaux:2002wm,Hotta:2008xt}.

Since   conserved charges depend on the coordinates, we need to
specify those concretely. Here, we will choose those as
$C_{\mp}=
\pm 1$ which gives us the standard metric form of BTZ black
holes.
That is to say, the background metric for ADT charge
computation,
which is $AdS_3$ space,  is taken in our coordinates as
\[
ds^2 = L^2\Big[ -r^2dt^2 + \frac{dr^2}{r^2} +
r^2d\theta^2\Big]\,.
\]
Then, masses and angular momenta of these black holes for the
Killing vectors $\xi_T = \frac{1}{L}\frac{\p}{\p t}$ and
$\xi_{R}= \pm \frac{\p}{\p \theta}$ are given by the so-called
ADT charge $Q^{\mu\nu}$ as\footnote{See~\cite{Nam:2010ma} for
our notation and some details about ADT formalism. Note also that   $\CE_{\mu\nu}$ expression only enters in the computation of the ADT charge  and  the so-called the correction term or hybrid formalism  by Cl\'{e}ment~\cite{Clement:2007} is not needed in this case since all our solutions satisfy the standard Brown-Henneaux boundary condtions.}
\bea M  &=& \frac{1}{4G}\sqrt{-\det g}~
Q^{rt}_{R}(\xi_T)\Big|_{r\rightarrow\infty} = \frac{1}{8G}
\Delta_0 \,, \\
J  &=& \frac{1}{4G}\sqrt{-\det g}~
Q^{rt}_{R}(\xi_R)\Big|_{r\rightarrow\infty} = \pm
\frac{L}{8G}\Delta_0\,. \nn
\eea
Therefore, masses and the total angular momenta satisfy the
extremal relation in these black holes as $ML =\pm J$.  This relation strongly suggests that any black hole solution  obtained from our reduced EOM is stable since the bound for angular momentum is saturated. 

Now, we argue that the inequality $W(\phi_H) \ge \CW(\phi_\infty) =2$ holds in general, which is an important ingredient to show the consistency with the holographic $c$-theorem.  On general ground, it is natural to think that masses of hairy black holes deformed from BTZ black holes by a scalar field are always greater than those of hairless BTZ black holes since the scalar hair produces additional positive contribution to masses. By accepting this assumption on mass inequality between extremal hairy black holes and extremal BTZ black holes, one can see that
\[ M(hairy) = \frac{\Delta_0}{8G} = \frac{r^2_H}{8G}\CW(\phi_H) \ge M(BTZ) = \frac{r^2_H}{4G}\,, \]
which imiplies that
\beq \CW(\phi_H) \ge\CW(\phi_\infty) =2\,. \eeq

The
Bekenstein-Hawking-Wald entropy of the above extremal black
holes
can be read from the Wald formula or the area law as
\beq \CS_{BHW} = \frac{A_H}{4G} = \frac{\pi
\bar{L}\bar{r}_H}{2G} = \frac{\pi L r_H}{2G} \,, \eeq
and the Hawking temperature of these black holes are always zero because of the extremality. This nature of zero Hawking temperature also indicates the stability of the hairy extremal black holes. 
The angular velocity of these black holes at the horizon $r_H$ is
given by
\beq
\Omega_H = \frac{1}{L}\left[C_{\mp} \mp
\frac{1}{r_H}e^{A(r_H)}\right]\,. \label{AngVel}
\eeq
Since the angular velocity of these black holes with $C_{\mp} =
\pm 1$ is given by $\Omega_H = \pm (1/L)$ 
and the Hawking temperature $T_H$ is zero,
one can check that the first law of black hole thermodynamics
is satisfied trivially. 
%
%

\subsection{Analytic solutions}

In this section, we present analytic solutions of our first order
reduced EOM.
According to the given setup, one can try to solve the last
differential equation (\ref{IntREOMDiff}) to obtain $r=r(\phi)$
for
the given superpotential $\CW$. Then, one can determine metric
functions $A$, $B$ and $C$ just by Eqs.~(\ref{MetricC}) and (\ref{MetricB}). In principle, this is the correct way to obtain solutions. However, it is not easy
to obtain {\it
analytic}
solutions in this way. Therefore, we take a slightly different
route: we will try to solve this equation by taking $r$ as an appropriate function of $\phi$. To find exact
solutions of the
above first order reduced EOM, let us try the simplest choice
for
$r^2$ just as  the asymptotic form itself:
\beq r^2 = \frac{4\omega_1}{(\phi - \phi_\infty)^2}\,. \eeq
%
%
%
%
%
%
Inserting this ansatz in the reduced differential
equation~(\ref{IntREOMDiff}), one can see that the
superpotential $\CW$ is given by
\bea
&&\CW = \alpha \Big[ 4 + (\phi - \phi_\infty)^2 \Big] + \beta\,
e^{(\phi-\phi_\infty)^2/4}\,,
\eea
where the constant $\alpha$ is related to $\Delta_0$ in
Eq.~(\ref{IntREOMDiff}) as $\Delta_0=4\alpha\omega_1$ and   $\beta$ is an arbitrary constant.
The metric functions
$A$, $B$ and $C$ are determined in terms of $\phi$ through
Eq.~(\ref{MetricB}), which can be converted to the functions of
$r$ as
\beq
e^{A} = r\Big[2\alpha\, e^{-\omega_1/r^2} +
\frac{\beta}{2}\Big]\,, \qquad
e^{B} = e^{-\omega_1/r^2} e^{-A}\,, \qquad e^{C} = \mp
\frac{1}{r}\, e^{A}\,,
\eeq
where  we have
rescaled the time coordinate as usual to absorb the integration
constant appropriately such that the asymptotic boundary
condition
on $A$ is  satisfied. Asymptotic boundary conditions on metric
functions also lead to $2\alpha + \beta/2 =1$. Now, let us impose the
existence of horizon through $e^{-B(r_H)}=0$, which leads to
\[
2\alpha + \frac{\beta}{2}\, e^{(\phi(r_H)-\phi_{\infty})^2/4}
=0\,.
\]
Then, one can obtain $\alpha$, $\beta$ in terms of
$r^2_H = 4\omega_1/(\phi_H-\phi_{\infty})^2$ as
\beq \alpha = \frac{1}{2}\frac{1}{1-e^{-\omega_1/r^2_H}}\,,
\qquad \beta = -\frac{2\,
e^{-\omega_1/r^2_H}}{1-e^{-\omega_1/r^2_H}}\,. \eeq
At last, one can see that black hole solutions given by $A(r),
B(r), C(r)$ and $\phi(r)$ are nothing but those
in Ref.~\cite{Hotta:2008xt}, which we have obtained in a different
way using the first order reduced EOM.

One can identify the near horizon geometry of these black holes
with the self-dual orbifold of $AdS_3$ with the radius $\bar{L}$
which is rescaled from the asymptotic radius $L$ as \beq \bar{L}
\equiv L\,  \bigg[\frac{1-e^{-\omega_1/r^2_H} }{\omega_1/ r^2_H}\bigg]\,.  \label{LIR1}  \eeq
To see this fact,  one may use a new radial coordinate $\rho$ defined by
\[
\rho \equiv 
e^{-\omega_1/r^2_H}\left[\frac{ 2\omega_1/r^2_H }{
1-e^{-\omega_1/r^2_H} } \right]^2\,(r-r_H)\,,
\]
and then one can explicitly check that the metric takes
the form given in Eq.(\ref{NHM}). This shows us that the above
hairy extremal AdS black holes interpolate between the
asymptotic $AdS_3$ with the radius $L$ and the self-dual
orbifold of $AdS_3$ with the radius $\bar{L}$.

Now, let us take a look at another analytic solution of our
first order equations. Under the successful reproduction of
known solutions, one may try another choice for $r^2$ as
\beq \ r^2 = \frac{4\omega_1}{\sinh^2(\phi - \phi_\infty)}\,.
\eeq
Following the same procedure in the above, one obtains
\bea 
W(\phi) &=& \alpha\Big[4 + \sinh ^2(\phi-\phi_{\infty}) \Big] +
\beta \Big[\cosh(\phi - \phi_{\infty}) \Big]^{1/2} \,, \nn \eea
where we have used the Eq.(\ref{IntREOMDiff}) to obtain this
result. The constant $\alpha$ is related to $\Delta_0(=-12\alpha
\omega_1)$, as in the previous case.
Then, metric functions $A$, $B$ and $C$ are  given by
\beq e^{A} = r \Big[ 2\alpha \Big( 1 +
\frac{4\omega_1}{r^2}\Big)^{\frac{3}{4}} +
\frac{\beta}{2}\Big]\,, \qquad e^{B} = \Big( 1 +
\frac{4\omega_1}{r^2}\Big)^{-\frac{1}{4}} e^{-A}\,, \qquad e^{C}
= \mp \frac{1}{r}\, e^{A}\,. \eeq
Asymptotic boundary conditions on $A$ and $B$ with the existence
of the horizon lead to
\beq \alpha =
-\frac{1}{2}\frac{1}{\big(1+\frac{4\omega_1}{r^2_H}\big)^{\frac{3}{4}}
-1} \,, \qquad \beta = \frac{2\big(1 +
\frac{4\omega_1}{r^2_H}\big)^\frac{3}{4} } { \big(1 +
\frac{4\omega_1}{r^2_H}\big)^\frac{3}{4} -1 }\,. \eeq
These are new  extremal AdS black hole solutions as far as the
authors know.  By introducing
\[
\rho \equiv \Big(1+
\frac{4\omega_1}{r^2_H}\Big)^{-\frac{1}{4}} \left[\frac{
6\omega_1/ r^2_H }{ \big(1+ 4\omega_1/r^2_H \big)^{3/4} -1}
\right]^2\, (r - r_H)  \,,
\]
one can show that the near horizon geometry is given by
self-dual orbifold of $AdS_3$ as the same form with
Eq.(\ref{NHM}) with $\bar{L}$ and $\bar{r}_{H}$ defined by
\beq \bar{L} \equiv L\, \left[\frac{ \big(1+ 4\omega_1/r^2_H
\big)^{3/4} -1} { 3\omega_1/ r^2_H }\right]\,, \qquad
\bar{r}_{H} \equiv r_{H}\, \Big[\frac{L}{\bar{L}} \Big]\,.
\label{LIR2} \eeq

As the final example,
we take the following ansatz
\beq r^2 = \frac{4\omega_1}{\sin^2(\phi - \phi_{\infty}) }\,.
\eeq
Then, the superpotential and metric functions are given by
\bea \CW &=& \alpha \Big[4+\sin^2(\phi - \phi_{\infty}) \Big] +
\beta \cos^{-\frac{1}{2}}(\phi - \phi_{\infty})\,, \\
e^{A} &=& r \bigg[ 2\alpha \Big( 1-
\frac{4\omega_1}{r^2}\Big)^{5/4} +\frac{\beta}{2}\bigg]\,,
\qquad
e^{B} = \Big( 1- \frac{4\omega_1}{r^2}\Big)^{1/4}\, e^{-A}\,,
\qquad e^{C} = \mp \frac{1}{r}\, e^{A}\,. \nn
\eea
and the constants $\alpha$ and $\beta$ are determined as
\beq \alpha =
\frac{1/2}{1-\big(1-\frac{4\omega_1}{r^2_H}\big)^{5/4} }\,,
\qquad \beta = -\frac{2\big(1-\frac{4\omega_1}{r^2_H}\big)^{5/4}
}{1-\big(1- \frac{4\omega_1}{r^2_H}\big)^{5/4}}\,. \eeq
Using the radial coordinate $\rho$  in this case  as
\[ \rho \equiv  \Big(1 -
\frac{4\omega_1}{r^2_H}\Big)^{\frac{1}{4}} \left[\frac{
10\omega_1/ r^2_H } {1- \big(1- 4\omega_1/r^2_H \big)^{5/4} }
\right]^2\,(r - r_H)\,,
\]
one can check that the near horizon geometry is once again given
by self-dual orbifold of $AdS_3$ in Eq.~(\ref{NHM}) with the
radius $\bar{L}$ and $\bar{r}_H $
\beq \bar{L} \equiv L \left[\frac{ 1 - \big(1- 4\omega_1/r^2_H
\big)^{5/4} }{ 5\omega_1/r^2_H} \right]\,, \qquad \bar{r}_{H}
\equiv r_{H} \Big[ \frac{L}{\bar{L}}\Big]\,. \label{LIR3}\eeq

Some comments for the above black hole solutions are in order.
Firstly, one may note that there is a new free parameter in the
above solutions denoted as $\omega_1$ which is related to the
scalar
field value  at the horizon, $\phi_H$,  and does not exist in
extremal BTZ black holes. As is obvious from  our method,
this
parameter $\omega_1$ is also related to the coefficient of the
leading term in the superpotential $\CW$. Secondly, one may note that the above solutions are the
extension of
extremal BTZ black holes to hairy cases and reduce to extremal BTZ black hole
solutions when the scalar field is turned off.  To see this
explicitly, one should take $\omega_1\rightarrow 0$ with the
frozen
scalar field $\phi =\phi_\infty$. Then, all the above
expressions
of black hole solutions reduce to those of  extremal BTZ black
holes.  This reveals  that the presence of a scalar field  may
produce diverse hairy black hole solutions via a scalar potential, which reduce to the same BTZ black
holes when a scalar field is turned off.  This point will  also  be   important to
understand
the nature of black hole solutions in NMG, which are presented  only in the perturbative form
in
the next section.

One may wonder about using the so-called Fefferman-Graham
coordinates in this case. That is to say,
 a new radial coordinate $\eta$  may be introduced  by
\[
   d\eta = e^{B(r)}dr\,.
\]
This gives us the so-called FG coordinates useful in later
sections and
corresponds to taking  the following form of the metric ansatz%
\beq
ds^2 = L^2 \Big[ -e^{2A(\eta)} dt^2 + d\eta^2
+e^{2R(\eta)}\Big(d\theta + e^{C(\eta)} dt\Big)^2
 \Big] \,. \label{FG}
\eeq
In these coordinates, most of the first order  reduced EOM in the
AdS-Schwarzschild coordinates remain as first order differential
equations\footnote{As in the AdS-Schwarzschild coordinates, we
take
the integration constant for $C$ such as $C=A-R$.} (recall that
$
r\equiv e^{R(\eta)}$)
\bea
\dot{\phi} = - \p_{\phi} \CW\,, \qquad \dot{A} + \dot{R} = \CW
\,, \qquad \dot{C} = \dot{A}-\dot{R} \,, \label{FGREOM}
\eea
where the dot  denotes the
differentiation
with respect to the radial coordinate $\eta$. 
However, the first order differential equation for $B(r)$ is
transformed to the second order one for $R(\eta)$ (or for
$\lambda \equiv e^{2R}$) as
\[
\ddot{\lambda} - \dot{\phi}(\p_{\phi}\CW) \lambda -\CW
\dot{\lambda}=0\,.
\]
Note that this
second
order differential equations is equivalent to $E_{\theta
\theta}
=0$ in these coordinates and corresponds to Eq.~(\ref{REOMDiff})
in
the $(r,t,\theta)$ coordinates. This means that the reduced EOM may
be
taken by the first order equations for $\phi$, $A+R$ and $C$
given
in Eq.~(\ref{FGREOM}), together with $E_{\theta \theta} =0$.
Interestingly, the second differential equation for $R$ (or
$E_{\theta \theta}=0$) can be integrated into the first order
form as
\[
\lambda  \CW  - \dot{\lambda}  = constant  = \Delta_0\,,\]
which corresponds to the Eq.~(\ref{IntREOMDiff}) in $(r,t,\theta)$ coordinates. It is also interesting
to note that
this integrated first order equation is automatically satisfied for domain wall
solutions which are included in our reduced EOM  as the special
case  given by
\beq
C=0\,, \qquad {\dot A}={\dot R} = \half \CW\,, \qquad
\Delta_0=0\,. \label{DWlimit} \eeq
This is consistent with our interpretation of  the constant
$\Delta_0$ as related to the near horizon of black holes, which
should be absent in domain walls. Though these coordinates cover only outside the horizon and do not
seem so useful in obtaining 
analytic solutions of AdS black holes in Einstein gravity, those simplify some computations and turn
out to be
particularly
useful
in NMG and in the holographic $c$-theorem, which are presented in
next
sections.

\section{Extremal Black Hole Solutions in NMG}
In this section we obtain lower order reduced EOM  in NMG and
then perturbative black hole solutions in NMG with an interacting
scalar field, which reduce to extremal BTZ black holes when
the scalar field is turned off.  In this section, we will confine ourselves to the two derivative theory for the scalar field, since the higher derivative terms for scalar field  lead to ghost instability and so they  are more  difficult to be analyzed.  One may  think that the higher derivative terms for  gravity is also problematic. However, as was analyzed in several works~\cite{Myers:2010xs}\cite{Sinha:2010ai}\cite{Camara:2010cd}, higher derivative terms for gravity  on an appropriate background can be treated effectively just like two derivative theory.  In the following we will follow this approach and consider   only  two derivative terms for the scalar field. 

\subsection{New massive gravity with a scalar field}
New massive gravity(NMG) is a three-dimensional higher curvature
gravity  introduced 
as the
covariant completion of Pauli-Fierz massive graviton
theory~\cite{Bergshoeff:2009hq}. Later it was recognized that
NMG
is more or less the unique extension of Einstein gravity
consistent
with the holographic
c-theorem~\cite{Myers:2010xs,Sinha:2010ai}.
In our convention, the Lagrangian of NMG with a scalar field is
given by
\begin{equation}\label{NMG}
S =\frac{1}{16\pi G}\int d^3x\sqrt{-g}\bigg[ \sigma R +
\frac{1}{m^2}\CK -\half \p_{\mu}\phi \p^{\mu}\phi -V(\phi)
\bigg]\,,
\end{equation}
where $\sigma$ takes $1$ or $-1$. The parameter $m^2$ can take
positive or negative values, and $\CK$ is a specific
combination of scalar curvature square  and Ricci tensor square
defined by \beq
 \CK = R_{\mu\nu}R^{\mu\nu} -\frac{3}{8}R^2\,.
\eeq
The equations
of motion(EOM)  of NMG are given by
\begin{eqnarray}\label{eomNMG}
0 = E_{\mu\nu} &\equiv & \CE_{\mu\nu} -  T_{\mu\nu}\,,
\end{eqnarray}
where
\begin{equation}
 \CE_{\mu\nu} \equiv \sigma \Big( R_{\mu\nu} -\half  R g_{\mu\nu}  \Big) +{1 \over {2 m^2}} \CK_{\mu \nu} \,, \quad T_{\mu\nu} \equiv   \half \p_{\mu}\phi \p_{\nu}
\phi  -\half g_{\mu\nu} \Big[ \half\p_{\alpha}\phi\p^{\alpha}\phi +   V(\phi)\Big] \nn \,,
\end{equation}
and ${\cal K}_{\mu\nu}$, using $\CD_{\mu}$ as a covariant
derivative with respect to $g_{\mu\nu}$, is defined by
\begin{equation}
\CK_{\mu\nu} =
g_{\mu\nu}\Big(3R_{\alpha\beta}R^{\alpha\beta}-\frac{13}{8}R^2\Big)
+ \frac{9}{2}RR_{\mu\nu} -8R_{\mu\alpha}R^{\alpha}_{\nu}+
\half\Big(4\CD^2R_{\mu\nu}-\CD_{\mu}\CD_{\nu}R
-g_{\mu\nu}\CD^2R\Big)\,. \label{Ktensor} \nn
\end{equation}
The equation of motion for the scalar field $\phi$ takes the
same form with Einstein gravity given in~Eq.(\ref{ScEOM}). 
We will focus on $\sigma=1$ case  and we set $\sigma=1$ in the following.  In order for the positive central charge of dual CFT with the  truncation of ghost modes~\cite{Maldacena:2011mk}, we also focus on  the positive $m^2$.

\subsection{Black hole solutions in new massive gravity}
As was done in the domain walls in NMG~\cite{Camara:2010cd}, let
us introduce the superpotential in NMG such that the scalar
potential is given by
\beq V(\phi) = \frac{1}{2L^2} \Big(\frac{\p \CW}{\p
\phi}\Big)^2\Big[1 - \frac{1}{8m^2L^2}\CW^2\Big]^2 -
\frac{1}{2L^2} \CW^2\Big[1 - \frac{1}{16m^2L^2}\CW^2\Big]\,.
\eeq
This generalized form of potential in terms of superpotential was first considered by Low and Zee in the context of scalar field coupled to higher derivative gravity~\cite{Low:2000pq}.
Motivated by results in Einstein gravity as given in
Eq.(\ref{FGREOM}),
 let us take the first order equations for  $A(\eta)$ and $C(\eta)$
\bea
\dot{\phi} = - \frac{\p \CW}{\p \phi}\Big[1 -
\frac{1}{8m^2L^2}\CW^2\Big]\,, \qquad \dot{A} + \dot{R} = \CW
\,, \qquad \dot{C} = \dot{A}-\dot{R} \,. \label{FGREOMNMG}
\eea
As in Einstein gravity, the last equation for $C$ can be
trivially integrated and may be omitted in the following.

One can check that these equations solve scalar EOM,
$E_{\phi}=0$ and metric EOM, $E_{\mu\nu}=0$ except $E_{\theta
\theta}=0$, even in NMG. Explicitly in these coordinates, the
scalar EOM and metric EOM except $E_{\theta \theta}=0$ can be
shown to be satisfied as follows:
\bea E_{\phi} &=& \ddot{\phi} +\CW \dot{\phi} - L^2\p_{\phi}V
=0\,, \\
-E_{\eta\eta} &=& L^2V + \half
\CW^2\Big[1-\frac{1}{16m^2L^2}\CW^2\Big] + \dot{\phi} \,
\p_{\phi}\CW \Big[1-\frac{1}{8m^2L^2}\CW^2\Big] + \half
\dot{\phi}^2 =0 \,, \nn \\
-e^{-A-R}E_{t\theta} &=& L^2 V+\frac{1}{2} \CW^2
\Big[1-\frac{1}{16 m^2 L^2}\CW^2\Big]
+\frac{1}{2} \dot{\phi} \, \p_{\phi}\CW \Big[1-\frac{1}{8 m^2
L^2}\CW^2\Big] =0\,. \nn \eea
%
%
%
Using $\lambda \equiv e^{2R}$, one can represent
$E_{\theta\theta}$ as
\bea
-2E_{\theta \theta} &=& \frac{1}{m^2L^2} \Big(\ddddot{\lambda}
-2\dddot{\lambda}\, \CW\Big) + \ddot{\lambda} \left[1+
\frac{9}{2m^2L^2}\Big(\frac{1}{4}\CW^2 - \dot{\phi}\p_{\phi} \CW
\Big) \right] \\
&& + \dot{\lambda}\, H_1(\CW,\p_{\phi}\CW, \cdots) + \lambda\,
H_2(\CW,\p_{\phi}\CW, \cdots) \,, \nn
\eea
%
%
where $H_1$ and $H_2$ are some functions 
of $\CW$ and its derivative with respect to $\phi$ given by
\begin{eqnarray}
H_1 &=& -{\CW  \over 2}\bigg(1+\frac{\CW^2}{8 m^2 L^2}\bigg) -{1 \over {8  m^2 L^2}} \left[ {14 \left(  {\dot \phi}^2 \, \p^2_{\phi}\CW +  {\ddot \phi} \,\p_{\phi}\CW  \right)} - {15 \, \CW \, {\dot \phi} \, \p_{\phi} \,\CW  } \right] \nn \\
H_2 &=& - L^2 V - {\CW^2 \over 2} \bigg(1-\frac{\CW^2}{16 m^2 L^2 }\bigg) - {   {\dot \phi}^2 \over {4 m^2  L^2}  }{\left(3 \left( \p_{\phi}\CW \right)^2+3 \CW \, \p^2_{\phi}  \, \CW - 2  \, {\dot \phi}  \,  \p^3_{\phi}  \, \CW \right)
 } \nn \\
&& +{1 \over {4 m^2  L^2}} \left[{\left( \CW^2 \, \p_{\phi} \CW + 6  \, \p^2_{\phi}\CW  \, {\ddot \phi} \right)  {\dot \phi}+\p_{\phi}\CW \left(3  \, \CW  \, {\ddot \phi} -2  {\dddot \phi} \right)} \right] \nn \,.
\end{eqnarray}
Now, one can see that the equation $E_{\theta \theta} =0$ can be
integrated as
\bea \tilde{\Delta}_0 
= \frac{1}{m^2L^2}\Big(\ddot{\Psi} - \CW\dot{\Psi}\Big) + \Big[1
+ \frac{1}{8m^2L^2}\Big(\CW^2 - 4\dot{\CW}\Big) \Big]\Psi\,,
\label{NMGREOMDiff}
\eea
where $\tilde{\Delta}_0$ denotes the integration constant and
$\Psi$ is defined as
\[
\Psi \equiv \lambda \CW - \dot{\lambda}\,.
\]
The physical meaning of the constant $\tilde{\Delta}_0$ turns
out to be similar to that of $\Delta_0$ in Einstein gravity.
That is to say, it is related to the conserved charges of black
holes and their near horizon geometry, which will be shown in
the below. Note also that we have reduced the fourth order EOM
effectively to the second order one.

By transforming $(\eta, t, \theta)$ coordinates to
$(\phi,t,\theta)$ coordinates, (which corresponds to taking
$\phi$ as the radial coordinate instead of $\eta$) as in
Einstein gravity, which corresponds to the following change of
variables
\beq
\frac{\p}{\p \eta} = \dot{\phi}~\p_{\phi} =
-(\p_{\phi}\CW)\Big[1 - \frac{1}{8m^2L^2}\CW^2\Big]\p_{\phi}\,,
\eeq
where we have used the reduced EOM for $\phi$ in the second
equality.
Through this transformation,    $\Psi$ is represented   as
\beq \Psi = r^2\CW + (\p_{\phi}r^2)(\p_{\phi}\CW)\Big[1 -
\frac{1}{8m^2L^2}\CW^2\Big]\,, \eeq
which should satisfy the differential
equation~(\ref{NMGREOMDiff}). When $\Psi$ is obtained,
 metric functions $A$ and $B$ can be given by
\bea A &=& -\ln r +\int^{\phi}d\phi'\, e^B\,\CW\, \p_{\phi'}r\,, \label{NMGMetric}
\\
e^{-B} &=& -(\p_{\phi}r)(\p_{\phi}\CW)\Big[1 -
\frac{1}{8m^2L^2}\CW^2\Big]= \frac{r}{2}\Big[\CW -
\frac{1}{r^2}\Psi\Big]\,. \nn \eea
These expressions for metric functions in $(r,t,\theta)$
coordinates come from the first order reduced EOM for $\phi$ and
$ A+ R$ in~(\ref{FGREOMNMG}), through the substitution of $\p/\p
\eta$ and $R$ by $e^{-B}\p/\p r$ and $\ln r$.

The differential equation for $\Psi$ is a nonlinear inhomogeneous
equation. It is not easy to obtain its solution analytically
except a trivial case.  Therefore, we try to obtain asymptotic
series form of solutions in NMG case, which might be
sufficiently
illuminating for discussion of holographic c-theorem in this
case.
As alluded in the previous section, the scalar hairy black holes with asymptotic AdS space and with
the near
horizon AdS space
would
correspond to the class of black holes which reduce to extremal
BTZ
black holes even in NMG. Before doing these perturbative
analysis,
let us consider the cases which allow analytic results.

Firstly, as in the Einstein gravity case, domain wall solutions
correspond to
\[
C=0\,, \qquad \dot{A}+\dot{R} =\CW\,, \qquad \Psi =0\,, \qquad
\tilde{\Delta}_0=0\,,\]
which allow analytic results and were studied
in Ref.~\cite{Camara:2010cd}.


Secondly, as a trivial example, let us check that extremal BTZ
black
holes are solutions of the above differential equation of
$\Psi$. By
taking $\CW = 2$, the first order reduced EOM for the scalar
field
$\phi$,  and metric function $\dot{A}+\dot{R}$,  lead to
\beq V = - \frac{2}{\ell^2}\,, \qquad \dot{\phi} = 0\,, \qquad
\dot{A} + \dot{R} = 2\,, \eeq
where $\ell$ is defined by
\[
\frac{1}{\ell^2} \equiv \frac{1}{L^2}\Big[1 -
\frac{1}{4m^2L^2}\Big] \,. \]
The second order differential equation for $\Psi$ gives us  a
constant  $\Psi$    as
\beq
\Psi = 2\lambda - \dot{\lambda} = 2r^2_H\,, \label{NMGBTZ}\eeq
where we have introduced the constant $r_H$ as $2 r^2_H \equiv
\tilde{\Delta}_0 [1+1/2m^2L^2]^{-1}$.
This gives us $e^{2R} = e^{2\eta} + r^2_H$, which corresponds to
the well-known extremal BTZ black hole solutions in NMG with the
horizon radius $r_H$.

To see nontrivial solutions one may try to solve the above
equations for a given superpotential.
However, one can see that the resulting solution for $\Psi$ is
given by a complicated function. Furthermore, it is not easy to
obtain analytic form of metric and scalar field in this way.  As in Einstein gravity, we perform the perturbative calculation at the asymptotic infinity and on the near horizon. This analysis already reveals important features of black hole solutions and is sufficient to verify that those black holes are extremal ones. 

Since the methodology is completely identical with Einstein gravity case, we briefly present  the intermediate steps. In summary, let us consider the following asymptotic expansions for metric variables, the superpotential and the scalar field:
\bea
A(r) &=& \ln r + \tilde{a}_1 r^{-2} + \cdots\,, \qquad B(r) = -\ln r +
\tilde{b}_1 r^{-2} + \cdots\,, \\
\CW &=& 2 + \frac{2\tilde{\omega}_1}{r^2} + \cdots = 2 +
\frac{1}{2q}(\phi-\phi_{\infty})^2
                + \cdots   \,,   \nn \\
\phi(r) &=& \phi_{\infty} + \frac{\tilde{\phi}_1}{r} + \cdots\,,
\nn
\eea
where $q$ denotes 
\[
  q \equiv 1 - \frac{1}{2m^2L^2}\,.
\]
It turns out that
$\tilde{a}_1$ and $\tilde{b}_1$ satisfy
$\tilde{a}_1+\tilde{b}_1 = - \tilde{\omega}_1$ and
$\tilde{\phi}^2_1 = 4q\tilde{\omega}_1$.  
By using the expansion of
$\Psi$ in terms of $r$ as
\beq \Psi = \Psi_0 + \frac{\Psi_1}{r^2} +\cdots\,, \eeq
and by solving the Eq.~(\ref{NMGREOMDiff})
perturbatively in terms of $r$, one obtains the following 
\[
 \Psi _0 \Big[1+\frac{1}{2m^2L^2}\Big] = \tilde{\Delta}_0\,.
\]
One can also  obtain  through Eq.~(\ref{NMGMetric})
\beq \tilde{a}_1 = -\half \tilde{\Delta}_0
\Big[1+\frac{1}{2m^2L^2}\Big]^{-1}\,, \qquad \tilde{b}_1 = \half
\tilde{\Delta}_0 \Big[1+\frac{1}{2m^2L^2}\Big]^{-1} -
\tilde{\omega}_1\,. \eeq
Note that this takes the form of hairy deformation from BTZ
black
holes given in the Eq.~(\ref{NMGBTZ}). Masses and angular momenta
of
these black holes can be obtained by the ADT method. Using the
results
given in Ref.~\cite{Nam:2010ma}, one can see that%
\beq M = \frac{1}{8G}\tilde{\Delta}_0 \,, \qquad J = \pm
\frac{L}{8G}\tilde{\Delta}_0 \,,\eeq
which satisfy the extremal condition $ML = \pm J$, as in
Einstein gravity
(see Ref.~\cite{Nam:2010ma} for more details about ADT charges in NMG
and how these give the above results.). As in Einstein gravity, it is straightforward to argue in NMG that the inequality, $\CW(\phi_H) \ge \CW(\phi_{\infty}) $ holds in general from mass inequality between hairy deformed extremal BTZ black holes and hairless ones, $M(hair) \ge M(BTZ)$.

Now, let us consider the expansions on the near horizon. By doing the perturbative analysis on the near  horizon, one can see
that
\beq \tilde{\Delta}_0 =
r^2_H\CW(\phi_H)\Big[1+\frac{1}{8m^2L^2}\CW^2(\phi_H)\Big]\,.
\eeq
By expanding the radial coordinate $r$ and the superpotential
$\CW$
in terms of $\phi$ as
\bea      r&=& r_H +\tilde{h}_0 (\phi-\phi_H) + \cdots\,, \nn \\
\CW(\phi) &=& \CW(\phi_H) -
\frac{1}{2}\CW(\phi_H)\Big[1-\frac{1}{8m^2L^2}\CW^2(\phi_H)\Big]^{-1}\,
(\phi-\phi_H) + \cdots\,, \nn \eea
which is important to see the relation between $\tilde{\Delta}_0$
and $\CW(\phi_H)$, one can also obtain metric
functions as, through the perturbative analysis, 
\beq
           e^{A(r)} = \tilde{s}_0 \CW(\phi_H) (r-r_H) + \cdots\,, \qquad
e^{B(r)} =  \frac{1}{\CW(\phi_H) (r-r_H)} + \cdots\,, \eeq
where $\tilde{s}_0$ is a certain non-vanishing constant related to the specific black holes or the interval of integral. 
As in Einstein gravity, one can show
that
\beq \bar{L} = \frac{2L}{\CW(\phi_H)}\,, \label{NNGbarL}\eeq
and can see that the extremality condition~(\ref{ExtCond}) is
fulfilled. This result shows us that the black holes under the consideration are extremal ones, indeed.
We also obtain the same results through the same expansions with original EOM without using the superpotential $\CW$ (see appendix B.)

Through the above analysis on the near horizon geometry of our extremal black holes,  one can see that the Bekenstein-Hawking-Wald entropy of these extremal
black holes in NMG  are given by
\beq \CS_{BHW} = \frac{A_H}{4G}\Big[1 +\frac{1}{2m^2\bar{L}^2}\Big] =  \frac{A_H}{4G}\Big[1 +\frac{1}{8m^2L^2}\CW^2(\phi_H) \Big]\,,
\qquad A_H \equiv 2\pi L r_H\,. \eeq
One can also verify that these are consistent with the first law of
black hole thermodynamics trivially as in Einstein gravity.
%
%

Let us consider the domain walls in this perturbative approach. By
taking the asymptotic form of the superpotential $\CW$ in the
same form with that in Einstein gravity as
\[
 \CW = 2 +\half (\phi-\phi_\infty)^2 + \cdots\,,
\]
one can obtain
asymptotic expansion of various variables in NMG    as
\bea A(r) &=& \ln r + \tilde{a}_1 r^{-2q} + \cdots\,, \qquad B(r) = -\ln r +
\tilde{b}_1 r^{-2q} + \cdots\,, \\
\CW &=& 2 + \frac{2\tilde{\omega}_1}{r^{2q}} + \cdots\,, \qquad \qquad \phi(r) = \phi_{\infty} +
\frac{\tilde{\phi}_1}{r^q} +
\cdots\,, \nn 
  \eea
where $\tilde{a}_1$ and $\tilde{b}_1$ satisfy $q\tilde{a}_1 + \tilde{b}_1 = -\omega_1$ and
$\tilde{\phi}^2_1
= 4\tilde{\omega}_1$. For a generic expansion of $\Psi$ in terms of
$\phi$,
one can see that all the coefficients of $\Psi$ vanish by
solving
the Eq.~(\ref{NMGREOMDiff}) perturbatively, {\it i.e.}
\beq \Psi =  \tilde{\Delta}_0 =  0\,, \eeq
which corresponds to the domain wall case. This short
computation
shows us that the asymptotic form of the superpotential for
domain wall solutions in NMG should be taken differently from those of black holes  and partially  explains to us why the reduced EOM for domain walls are different from those for black holes.

\section{Holographic C-Theorem}

In this section we consider the holographic $c$-theorem in the
context
of extremal AdS black holes in three dimensional Einstein
gravity
and in NMG.  By constructing central charge flow functions
holographically, one finds  some non-trivial checks of the
consistency between central charge expressions and parameters in extremal AdS black hole solutions.

A central charge flow function of the boundary field theory dual
to Einstein gravity may be introduced as
\beq C(\phi) = \frac{3L}{G}\frac{1}{\CW(\phi)}\,,\eeq
which gives us the central charge values on the asymptotic AdS
space and on the near horizon geometry as
\bea C(\phi\rightarrow \phi_{\infty}) &=& c_{UV} =
\frac{3L}{G}\frac{1}{\CW(\phi_\infty)} = \frac{3L}{2G}\,, \\
C(\phi\rightarrow \phi_{H}) &=& c_{IR} =
\frac{3L}{G}\frac{1}{\CW(\phi_H)} \equiv \frac{3L_{IR}}{2G}\,,
\nn
\eea
where we have introduced a IR scale $L_{IR}\equiv
2L/\CW(\phi_H)$.
Note that the superpotential value at the horizon,
$\CW(\phi_H)$,
is always greater than its asymptotic value
$\CW(\phi_{\infty})=2$,
which implies that  $c_{UV} \ge c_{IR}$. Moreover, $c_{UV}$ takes
the
standard value of two-dimensional boundary field theory dual to
$AdS_3$ with the radius $L$, which can be obtained by
Brown-Henneaux method~\cite{Brown:1986nw} or the standard
AdS/CFT
dictionary~\cite{Brown:1992br,Balasubramanian:1999re,Henningson:1998gx}.
Furthermore, in the domain wall limit, Eq.~(\ref{DWlimit}), the
central
charge flow function reduces to the well-known form  $C(\eta) =
3L/2G \cdot 1/A(\eta)$.

In the standard dictionary of  the AdS/CFT correspondence, one
identifies  the IR  scale $L_{IR}$ with the near horizon scale
$\bar{L}$ from the geometry. To verify the consistency of our
choice
of central charge flow functions, one needs to check that the 
central
charge flow functions reproduce the central charges of dual
conformal field theories even at the IR conformal point. Since
$\bar{L}$ is already determined by black hole parameters and
$L_{IR}$ is done by the superpotential values at the horizon,
two
results should be matched in order for the self-consistency of
our
construction. As was shown in Eq.~(\ref{barL}), the expressions
of
$\bar{L}$ is indeed identical with $L_{IR}$ as
\beq
\bar{L} = L_{IR} = \frac{2L}{\CW(\phi_H)}=
\frac{2r^2_HL}{\Delta_0}\,. \eeq
To see this explicitly for analytic solutions, one may note that for each ansatz of the radial coordinate $r$ in terms
of $\phi$
\[ r^2 = \frac{4\omega_1}{(\phi - \phi_\infty)^2}\,, \qquad
\frac{4\omega_1}{\sinh^2(\phi - \phi_{\infty}) }\,, \qquad
\frac{4\omega_1}{\sin^2(\phi - \phi_{\infty}) }\,, \]
 the superpotential values at the horizon, $\CW(\phi_H)$, are
given respectively by
\[ \CW(\phi_H) =
\frac{2\omega_1/r^2_H}{1-e^{-\omega_1/r^2_H}}\,, \qquad
\frac{6\omega_1/r^2_H}{\big(1+\frac{4\omega_1}{r^2_H}\big)^{3/4}
-1}\,, \qquad
\frac{10\omega_1/r^2_H}{1-\big(1-\frac{4\omega_1}{r^2_H}\big)^{5/4}}\,,
\]
which are consistent with the general analysis  as can be seen from   Eqs.(\ref{LIR1}), (\ref{LIR2}) and  (\ref{LIR3}).

Since we have verified that our central charge flow functions
lead
to the correct central charges at conformal end points, we turn
to
show  their monotonic properties.   According to the AdS/CFT
correspondence, it is well known that the scale in the RG flow
of
dual field theory corresponds to  the radial coordinate in the
so-called Fefferman-Graham coordinates in the gravity side. To
show
the monotonic property of the above central charge function
along
the RG flow in the dual field theory, one needs to consider the
derivative of the above central charge function with respect to
the
radial coordinate in the FG coordinates. The radial coordinate
$\eta$ introduced in Eq.~(\ref{FG})  forms the FG
coordinates together with $(\theta, t)$ in our case. Note that
$\eta\rightarrow \infty$ corresponds to the asymptotic infinity
(or
UV) and $\eta \rightarrow 0$ does to the near horizon (or IR).
Now,
one can see that
\beq 
\frac{d}{d\eta} C (\phi(\eta) ) =
-\frac{3L}{G}\frac{1}{\CW^2} \left(\p_{\phi}\, \CW \right)\, \dot{\phi} =
\frac{3L}{G}\frac{1}{\CW^2}(\p_{\phi}\CW)^2 \ge 0 \,,
\eeq
where we have used the first order equation for $\dot{\phi}$
which
is given in these coordinate as $\dot{\phi}= -\p_{\phi}\CW$.
This
result means that the central charge is always increased when
$\eta$
becomes increased (or when the energy scale is increased), and
this
can be regarded as the holographic construction of two
dimensional
$c$-theorem beyond the domain wall geometry.

Central charge flow function in NMG may be defined as
\beq C(\phi) = \frac{3L}{G}\frac{1}{\CW(\phi)} \Big[1 +
\frac{1}{8m^2L^2}\CW(\phi)^2\Big]\,,\eeq
which gives us central charge values at UV and IR as
\bea C(\phi_{\infty}) &=& c_{UV} = \frac{3L}{2G} \Big[1 +
\frac{1}{2m^2L^2}\Big]\,, \\
C(\phi_{H}) &=& c_{IR} = \frac{3L}{G} \frac{1}{\CW(\phi_{H})}
\bigg[1 + \frac{1}{8m^2L^2}\CW^2(\phi_{H}) \bigg] \equiv
\frac{3L_{IR}}{2G} \Big[1 + \frac{1}{2m^2L^2_{IR}}\Big]\,. \nn
\eea
where we have introduced $L_{IR} = 2L/\CW(\phi_H)$ as in the
case of Einstein gravity.
One may note that, on the contrary to Einstein gravity, $c_{UV}$
and $c_{IR}$ are not proportional to the cosmological constants
at the asymptotic infinity and at the near horizon.
It is quite useful to check $\bar{L} = L_{IR}$ for the whole
consistency of our results.
Indeed, one can see that this is the case by the asymptotic
analysis given in the previous section as was shown in
Eq.~(\ref{NNGbarL}).

The monotonic property of the chosen central charge flow
functions is anticipated in NMG. Using the first order equation
for scalar field given in Eq.~(\ref{FGREOMNMG}), one can verify
this anticipation as
\beq \dot{C} = -\frac{3L}{G}\frac{1}{\CW^2} \left(\p_{\phi}\CW \right) \Big[1
- \frac{1}{8m^2L^2}\CW^2\Big]~ \dot{\phi} =
\frac{3L}{G}\frac{1}{\CW^2} (\p_{\phi}\CW)^2 \Big[1 -
\frac{1}{8m^2L^2}\CW^2\Big]^2 \ge 0\,, \eeq
which represents the holographic construction of $c$-theorem
dual to extremal black holes in NMG.

Now, we give some comments about the relation between
holographic
$c$-theorem and  null energy condition on matters.  For domain
walls, one usually choose  a central charge flow function  as a
function of metric variables and then relate directly the derivative of central charge flow
functions, through
EOM, to
the null energy condition on matters.  On the contrary, the
connection between the holographic $c$-theorem and null energy
condition
is more or less indirect in our case, because the chosen central
charge function depends on superpotential not metric variables.
By
taking  null vectors on our geometry in  FG coordinates  as $
{\CN}_{\pm} = (N^{t}_{\pm}, N^{\eta}_{\pm}, N^{\theta}_{\pm} ) =(1/L)\big(\pm 1, ~e^{-R}, -e^{-A}
\big)$, one may
check that
null
energy condition on a scalar field is satisfied   as
\[
T_{\mu\nu}\CN^{\mu}_{\pm} \CN^{\nu}_{\pm} = \frac{1}{L^2}\,
\dot{\phi}^2 \ge 0\,.
\]
Only after using reduced EOM and not the original EOM, one may relate
$\dot{C}$ to $T_{\mu\nu}\CN^{\mu}_{\pm} \CN^{\nu}_{\pm}$. But
this
is somewhat indirect  connection to the null energy condition compared to the case of domain walls, which
use only the original EOM. This indicates that explicit
solutions
of EOM may be needed to verify the holographic $c$-theorem for more complicated geometry with two
asymptotic AdS
spaces as the
reduced
first order EOM are necessary to show those theorems in our
black
hole cases. 

In connection with the holographic $c$-theorem, it is
interesting to
consider the entropy of boundary dual CFT by Cardy formula.
Since
the bulk contains two $AdS_3$ spaces at the asymptotic infinity
and
at the near horizon, there are two entropy functions, $S_{UV}$
and
$S_{IR}$. One of the natural questions about these entropies is
that
there is any relation to the Bekenstein-Hawking-Wald entropy of
extremal black holes. The well-known relations between conserved
charges in the bulk gravity and energies in the dual CFT at the
asymptotic boundary  are given by~\cite{Strominger:1997eq}
\beq M = E_L + E_R\,, \qquad J = L(E_L-E_R)\,, \eeq
where $E_L/E_R$ are left/right energy in the dual CFT and
related to the so-called left/right temperatures as $E_{L/R} =
(\pi^2L/6)T^2_{L/R}$. The Cardy formula gives us the entropy
$S_{UV}$ in terms of these left/right temperatures as
\beq S_{UV} = \frac{\pi^2L}{3}\Big(c_LT_L +c_RT_R\Big)\,. \eeq
Note that $c = c_L = c_R$, since we are dealing with the parity
even
theories. The extremal condition in the black hole side means
that
one of the left and right energy should vanish in the dual CFT,
which can be deduced from the conserved charge relations.  For
definiteness, let us consider $ML=J$ case. After some
computation,
one can see that the entropy of dual CFT at the asymptotic
boundary
is not less than that the Bekenstein-Hawking-Wald entropy of
extremal black holes as\footnote{This fact is already observed
in Refs.~\cite{Hotta:2008xt,Carlip:1999cy} for particular black
holes in Einstein gravity.}
\bea S_{UV} &=& \frac{\pi^2L}{3}\, c_{UV}T_L = \pi
\sqrt{\frac{2}{3}c_{UV}J} \\
&=& \left\{ \ba{ll}
\frac{A_H}{4G}\Big[\frac{\CW(\phi_H)}{\CW(\phi_\infty)}\Big]^{1/2}
& ~~ {\rm Einstein} \\
\frac{A_H}{4G}\Big[1+\frac{1}{2m^2L^2}\Big]\Big[\frac{
\CW(\phi_H)}{\CW(\phi_\infty)}\frac{1+\frac{1}{8m^2L^2}\CW^2(\phi_H)}{1+\frac{1}{8m^2L^2}\CW^2(\phi_\infty)}
\Big]^{1/2} & ~~~
{\rm NMG}  \ea   \right.    \nn \\
&\ge& S_{BHW} \nn \,. \eea
Note that the inequality between $S_{UV}$ and $\CS_{BHW}$ is the direct consequence of the inequality between the
superpotential
values  $\CW(\phi_H) \ge \CW(\phi_\infty) =2$,  which can be verified explicitly in analytic solutions and was argued to be the case in general through mass inequalities between mass of hairy deformed extremal BTZ black holes and hairless ones.

In Einstein gravity, mass and angular momentum at the horizon,
$\bar{M}$ and $\bar{J}$ as quasi-local quantities were computed
and shown to satisfy $\bar{M}\bar{L} = \bar{J} =
J$~\cite{Hotta:2008xt} for specific extremal black holes. This
computation for quasi-local quantities on the near horizon
requires only the near horizon fall-off behavior of various
variables, our extremal black holes would satisfy the same
relations. One of the very interesting results in this
quasi-local computation is that the angular momentum is
invariant from the asymptotic infinity to the near horizon,
which was also observed in thermodynamic approach to black
holes  in a finite region~\cite{Brown:1994gs}. This invariance of angular momentum
in Einstein gravity leads to the entropy of dual CFT on the near
horizon through Cardy formula as
\beq S_{IR} = \pi \sqrt{\frac{2}{3}c_{IR}J} = \frac{A_H}{4G} =
\CS_{BHW}\,. \eeq
One may interpret this invariance of angular momentum along the
bulk radial direction in the dual CFT side as follows. Along the
RG flow in dual field theory, the running of central charge is
the consequence of the change of the AdS radius, while the bulk
gravitational constant, $G$, or the coupling $m^2$ of higher
curvature terms are invariant. From the bulk perspective in
our three-dimensional case, the AdS radius is the unique
candidate for the running variable, which is related to the
scalar field.  Based on   dimensional reasoning, one can see that the dual CFT energy
$E_{L/R}$,  which are computed from gravity, should scale inversely to the AdS radius  without anomalous scaling. Therefore,
the angular momentum is invariant with scaling of the AdS
radius.

By assuming the validity of the angular momentum invariance, one
can compute the entropy of dual CFT on the near horizon as
\beq S_{IR}  = \pi \sqrt{\frac{2}{3}c_{IR}J} =
\frac{A_H}{4G}\Big[1+ \frac{1}{8m^2L^2}\CW^2(\phi_H)\Big]  \,.  \eeq
It is very interesting to observe that the entropy of CFT on the near
horizon is also identical with the Bekenstein-Hawking-Wald entropy
of
extremal black holes,  as $S_{IR} =\CS_{BHW}$,   even in NMG\footnote{ Cautionary remarks for this statement: Since the original Cardy formula requires `effective central charge' $c_{eff}=c-24h_0$  ($h_0$: lowest conformal weight) not the original central charge $c$, as was reviewed in Ref.~\cite{Carlip:2005zn},  one should assume that $c$ is very large compared to $h_0$ or $h_0=0$ effectively,  for this matching.}.  In summary for all cases,
  one can see that the following inequalities always hold\footnote{We would like to thank an anonymous  referee for some improvements by pointing out some mistakes.}
\beq S_{UV} \ge S_{IR} = \CS_{BHW}\,. \eeq
%

\section{Conclusion}
We have discovered Bogomol'nyi type of lower order differential
equations for Einstein gravity and for NMG, which we called
reduced EOM, in the presence of a minimally coupled
scalar
field. More explicitly, the first order reduced EOM in Einstein
gravity is given by Eq.~(\ref{FirstDQ}) and the reduced EOM in
NMG
by Eqs.~(\ref{FGREOMNMG}) and (\ref{NMGREOMDiff}). Using these 
equations, we have obtained various
analytic hairy
black hole
solutions which include the previously known example as a
special
case. We also showed that all these solutions are consistent
with the 
holographic $c$-theorem.

The asymptotic space of all our solutions is AdS space with the
radius $L$ and the near horizon geometry is the so-called
self-dual
orbifold of $AdS_3$ with the radius $\bar{L}$. After showing
that
the simplest case of our black hole solutions corresponds to the
well-known extremal BTZ black holes, the extremality of our
black
hole solutions are shown explicitly. We also showed that   our
reduced EOM implies generically the extremality of any solution
of
these equations under some mild assumptions, the validity of the 
power series expansion. In Einstein
gravity we
have presented
several analytic solutions with some generic perturbative
treatment
and in NMG we have done some perturbative solutions. We have
also
identified various physical quantities of extremal black hole
solutions and shown that conserved charges of these black holes
are
related directly to the integration constant $\Delta_0$ and
$\tilde{\Delta}_0$ of the reduced EOM.

Motivated from the domain wall case, we have
proposed the holographic central charge flow functions and
verified that they coincide with central charges at the
conformal end points and that they also satisfy the anticipated
monotonic properties. We have also performed the consistency
checks on the central charge flow functions by showing that the
near horizon scale $\bar{L}$, which is read from the geometry,
can be identified with the IR scale $L_{IR}$ in the central
charge flow functions.

There are some future directions to pursue further. First of
all,
our reduced EOM  are just rewritten down in analogy with the
domain
wall case. It would be very interesting to derive our reduced
EOM  
by complete squaring of a certain energy functional or by using
fake
supersymmetry. It is notable that the second order original EOM 
in
Einstein gravity reduce to the first order ones while all of the reduced
EOM 
in NMG are not first order ones. This is clearly contrasted to the
domain
wall case which is also described by the first order ones even in
NMG.
It would be very interesting to reproduce conserved charges for
our
extremal black holes consistently in Einstein gravity and NMG by other methods. Another interesting direction is to verify the stability of our extremal black holes directly. Though we showed that all the black hole solutions from our reduced EOM are extremal and argued that they are all stable  by the generic thermodynamic consideration, we did not perform any  dynamical stability analysis through some perturbations.

 It is also interesting to prove the inequality $\CW(\phi_H) \ge \CW(\phi_{\infty})$ rigorously for any sensible black hole solutions, which is argued to be the case on physical ground.
It would be also interesting to verify or disprove our
conjecture about the invariance of angular momentum along RG
flow in general. In Einstein gravity, the invariance is rather
established in various cases. However, it is not checked
explicitly in other gravity theories like NMG.
It is also valuable to
find more solutions or to study the complete integrability of
our first order reduced EOM  in Einstein gravity.
Another intriguing direction is the study on couplings with
more than one scalar field. For instance, it might be possible
to obtain vortex black hole solutions when a complex scalar field
is coupled.
Finally, it would be very interesting to extend the
correspondence between DLCQ of two-dimensional field theory and
extremal black holes to our hairy cases.
\vskip 3cm



\centerline{\large \bf Acknowledgments} 
\vskip0.5cm

{S.N and S.H.Y were supported by the National
Research Foundation(NRF) of Korea grant funded by the Korea
government(MEST) through the Center for Quantum
Spacetime(CQUeST) of
Sogang University with grant number 2005-0049409. S.H.Y was supported
  by Basic Science Research Program through the NRF of Korea funded by the MEST(2012R1A1A2004410). S.H.Y  would  
like to thank Prof. Won Tae Kim for some useful discussions.    J.D.P was supported by a
grant from the Kyung Hee University in 2009(KHU-20110060) and Basic Science Research Program through the NRF of Korea  funded by the MEST(2012R1A1A2008020). 
S.N  was supported by Basic Science Research Program through the NRF of Korea funded by the MEST(No.2012-0001955).  Y.K  was  supported by the NRF of Korea grant funded by the Korean government(MEST) (NRF-2011-355-C00027). YK would  
like to thank Dr. Hyojoong Kim for some helpful discussions.  }

\newpage

\appendix
\centerline{\large \bf Appendix}

 \renewcommand{\theequation}{A.\arabic{equation}}
  \setcounter{equation}{0}
\section{Equations of motion in NMG for some coordinate systems}

In this section, we present EOM   in NMG only since EOM in Einstein gravity can be obtained by taking $m^2L^2\rightarrow \infty$ in the following expressions of EOM in NMG.

For our metric ansatz~(\ref{BTZ}) with the relation (\ref{MetricC}), we present the metric EOM in the following since the scalar EOM is already given in Eq.~(\ref{ScEOM}).  
\bea
0&=& E_{rr} = -L^2e^{2 B} V -\frac{1}{2r} \Big[r A'^2+\left(2-2 r
B'\right) A'+r \phi'^2-2 B'+2 r A''\Big]  + \frac{1}{2 r^2}  \\
&&\qquad +{e^{-2 B} \over {32 m^2 L^2 r^4 } } {(r A'+1)^2 \Big[ r
\big(A' (r A'-4 r B'+2)+4 r A''-4 B' \big)-3 \Big]}\,,  \nn \\
0&=& E_{\theta \theta} = -L^2r^2 V -\frac{1}{2} e^{-2 B} (r^2 A'^2-2
r B'+1 ) \nn \\
&&\qquad+\frac{e^{-4 B}}{32 m^2 L^2 r^2} \bigg[ A'^4 r^4-24 A''^2
r^4+96 B'^3 r^3-8 A'^3 \left(r B'-1\right) r^3+80 A'' r^2 \nn \\
&&\qquad +8 B'^2 \left(22 r^2 A''-13\right) r^2+40 B'' r^2-7 -64 A''
B'' r^4+88 A''' r^3 \nn \\
&&\qquad +16 A'''' r^4+16 B''' r^3 +2 A'^2 r^2 \left(4 A'' r^2+12 B''
r^2+66 B' r-36 B'^2 r^2-13\right) \nn \\
&&\qquad -4 B' r (24 A''' r^3+90 A'' r^2+28 B'' r^2+5 )   \nn  \\
&&\qquad -8 A' r^2 \Big(12 r^2 B'^3-44 r B'^2+ (21-15 A'' r^2-14 B''
r^2 ) B' \nn \\
&&\qquad  +r (13 A''+16 B''+3 r A'''+2 r B''')\Big)  \bigg] \,,   \nn  \\
0&=& E_{t \theta} =\mp \frac{1}{2} e^A \bigg[ e^{-2 B} \big( B'-A' (r
A'-r B'+2)-r A'' \big)-2 L^2 r V \bigg]  \nn \\
&&\qquad \mp {e^{A-4 B} \over {32 m^2 L^2 r^3} } (r A'+1)^2 {\bigg[ r
\big( A' (r A'-2 r B'+2 )-2 B'+2 r A'' \big) -1\bigg]}\,,  \nn
\eea
 where $'$ denotes  differentiation with respect to the radial coordinate $r$.

For the metric ansatz (\ref{FG}) with the relation $C= A-R$, which is written in the so-called FG coordinates, the scalar field equation and EOM  in
NMG are given by
\bea
0&=&E_{\phi} =  \frac{1}{L^2} \Big[ ({\dot A} +{\dot R}) {\dot \phi
}+{\ddot \phi} \Big] -\partial_{\phi} V\,, \\
0&=& E_{\eta \eta} =  - V L^2
 -\frac{1}{2} \bigg[ \left({\dot A}+{\dot R} \right)^2 +
{\dot \phi}^2 +2 \left({\ddot A}+{\ddot R} \right) \bigg] \nn \\
&&\qquad  +{1 \over {32 m^2 L^2 }} \left({\dot A}+{\dot R}\right)^2 {
\bigg[ \left({\dot A}+{\dot R}\right)^2+4 \left({\ddot A}+{\ddot
R}\right)\bigg]}\,,   \nn  
\eea
\bea
0&=& E_{\theta \theta} = -\frac{1}{2} e^{2 R}
\left(2 V L^2+{\dot A}^2+3 {\dot R}^2+2 {\ddot R}\right) \nn \\
&&\qquad + {e^{2 R} \over {32 m^2 L^2 }}
\bigg[ {\dot
A}^4+8 {\dot R}
{\dot A}^3+ {\dot A}^2 \left(8 {\ddot A}-54 {\dot R}^2 -28
{\ddot R}\right) -67
  {\dot R}^4   \nn    \\
&&\qquad +4 {\dot R}^2 \left(44 {\ddot A}-61 {\ddot R}\right) +8 {\dot
A} \left(16 {\dot R}^3-13 \left({\ddot A}-2 {\ddot R}\right)
{\dot R} -3 {\dddot A}+5 {\dddot R} \right) \nn \\
&&\qquad -8 \left(3 \left({\ddot A}-3 {\ddot R}\right) \left({\ddot
A}-{\dot R}\right)-2 {\ddddot A}+2 {\ddddot R} \right) +8 {\dot
R} \left(11 {\dddot A} -13 {\dddot R} \right)\bigg]\,, \nn
\\
0&=& E_{t \theta} =  -\frac{1}{2} e^{A+R} \left(2 V L^2+\left({\dot
A}+{\dot R}\right)^2+{\ddot A}+{\ddot R}\right)
\nn  \\
&&\qquad  +  {{e^{A+R} } \over {32 m^2 L^2
}} {\left({\dot A}+{\dot R}\right)^2 \bigg[ \left({\dot A}+{\dot
R}\right)^2+2 \left({\ddot A}+{\ddot R}\right)\bigg]}\,, \nn
\eea
 where $\dot{}$ denotes  differentiation with respect to the new radial coordinate $\eta$.

 \renewcommand{\theequation}{C.\arabic{equation}}
  \setcounter{equation}{0}
\section{Perturbative calulation in the original EOM}
In this section, we perform some perturbative calculations in the original EOM, which reveal that full EOM need to be organized in terms of a certain formal expansion parameter $\epsilon$.   This formal expansion parameter $\epsilon$ is just the book-keeping one  to retain the correct order in EOM and obtain consistent results, which should be set to be unity at the end of calculation.  Only the relevant orders of $\epsilon$ will be represented in the expansion expressions below. Furthermore, we do not use the trick exchanging the role of the radial coordinate and the scalar field $\phi$. All the variable are expanded in terms of the radial coordinate $r$.  Therefore, this calculation may be regarded as the independent check of our results given in the main text. In the following, we will use $(t,r,\theta)$ coordinate to show our results. 

At the asymptotic of extremal black holes,  we can consider following expansion of various
variables in NMG as
\bea 
A(r) &=& \ln r +\epsilon^2 \tilde{a}_1 r^{-2} + \cdots\,, \qquad B(r)
= -\ln r + \epsilon^2 \tilde{b}_1 r^{-2} + \cdots\,, \\ \phi(r) &=& \phi_{\infty} + \epsilon
\frac{\tilde{\phi}_1}{r} +
\cdots\,, \qquad 
V(\phi) =  - {2 \over L^2} \left( 1- {1 \over {4 m^2L^2}}
\right) -{1 \over {2 L^2}} (\phi
   -\phi_{\infty})^2+ \cdots\,,  \nn \eea
With the above expansion and the parameter $\epsilon$, the expressions for the scalar field equation and the metric EOM in NMG are obtained as
\begin{eqnarray}
E_{\phi} &=&\CO(\epsilon^{3}) + \cdots\,,  \qquad
E_{rr}  = \CO(\epsilon^{4})+ \cdots\,,  \nn \\
E_{\theta\theta} &=& \frac{1}{2} \left[4q
(\tilde{a}_1+\tilde{b}_1) + {\tilde{\phi}_1}^2 \right] \epsilon^2 +\CO(\epsilon^{4}) + \cdots\,, \nn \\
E_{t\theta} &=& \frac{1}{2} \left[4 q(\tilde{a}_1+\tilde{b}_1)
+ {\tilde{\phi}_1}^2 \right] \epsilon^2 +\CO(\epsilon^{4}) + \cdots\,, \nn
\end{eqnarray}
where $q \equiv 1 - 1/2m^2L^2$.

Through $0=E_{\phi}$ and $0=E_{\mu\nu}$,   we obtain the following relation
\begin{eqnarray}
q (\tilde{a}_1+\tilde{b}_1) =- \frac{1}{4}{\tilde{\phi}_1}^2\,,
\end{eqnarray}
which is consistent with the results from reduced EOM in NMG. However, as we have alluded to  in the main text, the reduced EOM can be integrated partially with the integration constant $\tilde{\Delta}_0$, the constants $\tilde{a}_1$ and $\tilde{b}_1$ can be determined completely in terms of $\tilde{\Delta}_0$ and $\tilde{\omega}_1$, while that is not possible in the perturbative calculation in the full EOM.

On the near horizon, we have the expansion of various variables as
\bea 
A(r) &=& \ln {a\,(r-r_H)} + \cdots\,, \qquad B(r) = - \ln
{b\,(r-r_H)} + \cdots\,, \\ 
\phi(r) &=& \phi_H +  \frac{\epsilon}{\tilde{h}_0} (r-r_H)+ \cdots\,, \nn \\ V(\phi)&=& -\frac{ {\CW(\phi_H) }^2 }{2
L^2}\bigg[{
\bigg(1-\frac{ {\CW(\phi_H) }^2}{16 m^2 L^2 }\bigg)} - 2(\phi -
 \phi_H)^2 \bigg]+ \cdots\,. \nn 
\eea
With the above expansion, the expressions for scalar field equation and the  metric EOM 
 can be written as
\begin{eqnarray}
E_{\phi} &=&\CO(\epsilon) + \cdots\,,  \nn \\
E_{rr} &=&{1 \over {32 b^2 m^2 L^2 }}{\left(b^2- {\CW(\phi_H)
}^2\right) \left(b^2-16 m^2 L^2
+ {\CW(\phi_H) }^2\right)} (r-r_H)^{-2} + \CO(\epsilon^{2}) +
\cdots\,, \nn \\
E_{\theta\theta} &=& { {r_H}^2 \over {32 m^2 L^2 }}{ \left(b^2-
{\CW(\phi_H) }^2\right) \left(b^2-16 m^2 L^2
+ {\CW(\phi_H) }^2\right)} +\CO(\epsilon^{2}) + \cdots\,, \nn \\
E_{t\theta} &=& { {a \, r_H} \over {32 m^2 L^2 }}{ \left(b^2-
{\CW(\phi_H) }^2\right) \left(b^2-16 m^2 L^2
+ {\CW(\phi_H) }^2\right)} (r-r_H)+\CO(\epsilon^{2}) + \cdots\,.
\nn
\end{eqnarray}
From $0=E_{\phi}$ and $0=E_{\mu\nu}$,    we obtain
\bea 
b^2={\CW(\phi_H)}^2\,,
\eea
which is also consistent with the results obtained from the reduced EOM. 
 Note that there is another  possibility such that $b^2-16 m^2 L^2
+ {\CW(\phi_H) }^2 =0$, which has no   Einstein limit and seems to indicate the existence of  solutions different from those for the  reduced EOM.


\newpage


\begin{thebibliography}{99}


\bibitem{Komargodski:2011vj}
  Z.~Komargodski and A.~Schwimmer,
{\it On renormalization group flows in four dimensions},
  JHEP {\bf 1112} (2011) 099
  [arXiv:1107.3987 [hep-th]].


  Z.~Komargodski,
  {\it The Constraints of Conformal Symmetry on RG Flows},
  JHEP {\bf 1207} (2012) 069
  [arXiv:1112.4538 [hep-th]].




\bibitem{Cardy:1988cwa}
  J.~L.~Cardy,
  {\it Is there a c theorem in four-dimensions?},
  Phys.\ Lett.\ B {\bf 215} (1988) 749.


  H.~Osborn,
 {\it Derivation of a four-domensional c theorem},
  Phys.\ Lett.\ B {\bf 222} (1989) 97.


\bibitem{Casini:2011kv}
  H.~Casini, M.~Huerta and R.~C.~Myers,
  {\it Towards a derivation of holographic entanglement entropy},
  JHEP {\bf 1105} (2011) 036
  [arXiv:1102.0440 [hep-th]];

   P.~Calabrese and J.~L.~Cardy,
{\it  Entanglement entropy and quantum field theory},
   J. Stat. Mech {\bf 0406} (2004) P06002
   [arXiv:hep-th/0405152].


  H.~Casini and M.~Huerta,
  {\it On the RG running of the entanglement entropy of a circle},
  Phys.\ Rev.\ D {\bf 85} (2012) 125016
  [arXiv:1202.5650 [hep-th]].



\bibitem{Jafferis:2010un}
  D.~L.~Jafferis,
  {\it The exact superconformal R-symmetry extremizes Z},
  JHEP {\bf 1205} (2012) 159
  [arXiv:1012.3210 [hep-th]].



  I.~R.~Klebanov, S.~S.~Pufu and B.~R.~Safdi,
  {\it F-Theorem without Supersymmetry},
  JHEP {\bf 1110} (2011) 038
  [arXiv:1105.4598 [hep-th]].



  C.~Closset, T.~T.~Dumitrescu, G.~Festuccia, Z.~Komargodski and N.~Seiberg,
{\it Contact Terms, Unitarity, and F-Maximization in Three-Dimensional Superconformal Theories}, 
  arXiv:1205.4142 [hep-th].





\bibitem{Freedman:1999gp}
  D.~Z.~Freedman, S.~S.~Gubser, K.~Pilch and N.~P.~Warner,
 {\it Renormalization group flows from holography supersymmetry and
a c theorem},
  Adv.\ Theor.\ Math.\ Phys.\  {\bf 3} (1999) 363
  [hep-th/9904017];


  L.~Girardello, M.~Petrini, M.~Porrati and A.~Zaffaroni,
{\it Novel local CFT and exact results on perturbations of N=4
superYang Mills from AdS dynamics},
  JHEP {\bf 9812} (1998) 022
  [hep-th/9810126];


  J.~Distler and F.~Zamora,
{\it Nonsupersymmetric conformal field theories from stable anti-de
Sitter spaces},
  Adv.\ Theor.\ Math.\ Phys.\  {\bf 2} (1999) 1405
  [hep-th/9810206].


\bibitem{de Boer:1999xf}
  J.~de Boer, E.~P.~Verlinde and H.~L.~Verlinde,
  {\it On the holographic renormalization group},
  JHEP {\bf 0008} (2000) 003
  [hep-th/9912012].


\bibitem{Myers:2010xs}
  R.~C.~Myers and A.~Sinha,
  {\it Seeing a c-theorem with holography},
  Phys.\ Rev.\ D {\bf 82} (2010) 046006
  [arXiv:1006.1263 [hep-th]];

  R.~C.~Myers and A.~Sinha,
  {\it Holographic c-theorems in arbitrary dimensions},
  JHEP {\bf 1101} (2011) 125
  [arXiv:1011.5819 [hep-th]];

  R.~C.~Myers and A.~Singh,
{\it Comments on holographic entanglement entropy and RG flows},  
JHEP {\bf 1204} (2012) 122
  [arXiv:1202.2068 [hep-th]].


\bibitem{Zamolodchikov:1986gt}
  A.~B.~Zamolodchikov,
 {\it Irreversibility of the flux of the renormalization group in
a 2D field theory},
  JETP Lett.\  {\bf 43} (1986) 730 ;
  Pisma Zh.\ Eksp.\ Teor.\ Fiz.\  {\bf 43} (1986) 565.


\bibitem{Skenderis:1999mm}
  K.~Skenderis and P.~K.~Townsend,
  {\it Gravitational stability and renormalization group flow},
  Phys.\ Lett.\ B {\bf 468} (1999) 46
  [hep-th/9909070].


\bibitem{DeWolfe:1999cp}
  O.~DeWolfe, D.~Z.~Freedman, S.~S.~Gubser and A.~Karch,
  {\it Modeling the fifth-dimension with scalars and gravity},
  Phys.\ Rev.\ D {\bf 62} (2000) 046008
  [hep-th/9909134].

\bibitem{Low:2000pq} 
  I.~Low and A.~Zee,
  {\it Naked singularity and Gauss-Bonnet term in brane world scenarios},
 Nucl.\ Phys.\ B {\bf 585}, 395 (2000) 
 [hep-th/0004124]. 



\bibitem{Henneaux:2002wm}
  M.~Henneaux, C.~Martinez, R.~Troncoso and J.~Zanelli,
{\it Black holes and asymptotics of 2+1 gravity coupled to a scalar
field},
  Phys.\ Rev.\ D {\bf 65} (2002) 104007
  [hep-th/0201170];

  C.~Martinez, R.~Troncoso and J.~Zanelli,
  {\it Exact black hole solution with a minimally coupled scalar field},
  Phys.\ Rev.\ D {\bf 70} (2004) 084035
  [hep-th/0406111].

\bibitem{Hotta:2008xt}
K.~Hotta, Y.~Hyakutake, T.~Kubota, T.~Nishinaka and H.~Tanida, 
{\it The CFT-interpolating black hole in three
  dimensions},
  JHEP {\bf 0901} (2009) 010
  [arXiv:0811.0910 [hep-th]].


\bibitem{Bergshoeff:2009hq}
  E.~A.~Bergshoeff , O.~Hohm and P.~K.~Townsend,
  {\it Massive gravity in three dimensions},
  {\PRL} {\bf 102}  (2009) 201301
  [arXiv:0901.1766 [hep-th]];

   E.~A.~Bergshoeff , O.~Hohm and P.~K.~Townsend,
   {\it More on massive 3D gravity},
   {\PR} D {\bf 79}  (2009) 124042
  [arXiv:0905.1259 [hep-th]];

  S.~Nam, J.~D.~Park and S.~H.~Yi,
{\it AdS black hole solutions in the extended new massive
gravity},
   {\JHEP}{\bf 1007} (2010) 058
  [arXiv:1005.1619 [hep-th]];

  I.~Gullu, T.~C.~Sisman and B.~Tekin,
  {\it Born-Infeld extension of new massive gravity},
   \CQG {\bf 27} (2010) 162001
  [arXiv:1003.3935 [hep-th]];

  R.~Jackiw, S.~Templeton and S.~Deser,
  {\it Three-dimensional massive gauge theories},
   \PRL {\bf48} (1982) 975;

  S.~Deser, R.~Jackiw and S.~Templeton,
  {\it Topologically massive gauge theories},
   {\it Ann. Phys.}  {\bf 140} (1982) 372
  [Erratum-ibid.\ 1988 {\bf 185} 406 , 1988 {\bf 281} 409];

  W.~Li, W.~Song and A.~Strominger,
  {\it Chiral gravity in three dimensions},
  JHEP {\bf 0804} (2008) 082
  [arXiv:0801.4566 [hep-th]].


\bibitem{Sinha:2010ai}
  A.~Sinha,
  {\it On the new massive gravity and AdS/CFT},
   {\JHEP}{\bf 1006} (2010) 061
  [arXiv:1003.0683 [hep-th]];

  A.~Sinha,
{\it On higher derivative gravity, $c$-theorems and cosmology}, Class.\ Quant.\ Grav.\ {\bf 28}
(2011) 085002
  [arXiv:1008.4315 [hep-th]].

\bibitem{Camara:2010cd}
  U.~d.~.Camara and G.~M.~Sotkov,
 {\it New massive gravity domain walls},
  Phys.\ Lett.\ B {\bf 694} (2010) 94
  [arXiv:1008.2553 [hep-th]];

  U.~Camara da Silva, C.~P.~Constantinidis, A.~L.~Alves Lima and G.~M.~Sotkov,
  {\it Domain walls in extended Lovelock gravity},
  JHEP {\bf 1204} (2012) 109
  [arXiv:1202.4682 [hep-th]].

  G.~M.~Sotkov and U.~Camara dS,
  {\it Holographic RG flows from quasi-topological gravity},
  arXiv:1207.0778 [hep-th].
 
\bibitem{Banados:1992wn}
  M.~Ba\~{n}ados, C.~Teitelboim and J.~Zanelli,
  {\it The Black hole in three-dimensional space-time},
  \PRL {\bf 69} (1992) 1849
  [arXiv:hep-th/9204099].

\bibitem{Brown:1986nw}
  J.~D.~Brown and M.~Henneaux,
{\it Central charges in the canonical realization of asymptotic
symmetries: An
  example from three-dimensional gravity},
 {Commun. Math. Phys.}  {\bf 104} (1986) 207.

\bibitem{Coussaert:1994tu}
  O.~Coussaert and M.~Henneaux,
{\it Selfdual solutions of (2+1) Einstein gravity with a negative
cosmological constant},
  In *Teitelboim, C. (ed.): The black hole* 25-39
  [hep-th/9407181].


\bibitem{Balasubramanian:2009bg}
V.~Balasubramanian, J.~de Boer, M.~M.~Sheikh-Jabbari and
J.~Simon,
{\it What is a chiral 2d CFT? And what does it have to do with
extremal black holes?},
  JHEP {\bf 1002} (2010) 017
  [arXiv:0906.3272 [hep-th]].




\bibitem{Maldacena:2011mk}
  J.~Maldacena,
  {\it Einstein Gravity from Conformal Gravity},
  arXiv:1105.5632 [hep-th].

  S.~-J.~Hyun, W.~-J.~Jang, J.~-H.~Jeong and S.~-H.~Yi,
  {\it Noncritical Einstein-Weyl Gravity and the AdS/CFT Correspondence},
  JHEP {\bf 1201} (2012) 054
  [arXiv:1111.1175 [hep-th]].
  Y.~Kwon, S.~Nam, J.~-D.~Park and S.~-H.~Yi,
  {\it AdS/BCFT Correspondence for Higher Curvature Gravity: An Example},
  JHEP {\bf 1206} (2012) 119
  [arXiv:1201.1988 [hep-th]].
  
\bibitem{Abbott:1981ff}
  L.~F.~Abbott and S.~Deser,
  {\it Stability of Gravity with a Cosmological Constant},
  Nucl.\ Phys.\ B {\bf 195} (1982) 76.
  
  
  S.~Deser and B.~Tekin,
 {\it Gravitational energy in quadratic curvature gravities},
  Phys.\ Rev.\ Lett.\  {\bf 89} (2002) 101101
  [hep-th/0205318].


  S.~Deser and B.~Tekin,
{\it Energy in generic higher curvature gravity theories}, 
  Phys.\ Rev.\ D {\bf 67} (2003) 084009
  [hep-th/0212292].






\bibitem{Nam:2010ma}
  S.~Nam, J.~D.~Park  and S.~H.~Yi,
{\it Mass and angular momentum of black holes in new massive
gravity},
   \PR D {\bf 82} (2010) 124049
  [arXiv:1009.1962 [hep-th]].
  

\bibitem{Clement:2007}
  A.~Bouchareb and G.~Cl\'{e}ment,
 {\it Black hole mass and angular momentum in topologically massive gravity},
  Class.\ Quant.\ Grav.\ {\bf 24} (2007) 5581
  [arXiv:0706.0263 [gr-qc]].

  G.~Cl\'{e}ment,
  {\it Warped AdS$_3$ black holes in new massive gravity},
  Class.\ Quant.\ Grav.\  {\bf 26} (2009) 105015 
  [arXiv:0902.4634 [hep-th]].
  
  

\bibitem{Brown:1992br}
  J.~D.~Brown and J.~W.~York,
{\it Quasilocal energy and conserved charges derived from the
gravitational
  action},
   \PR  D {\bf 47} (1993) 1407
  [arXiv:9209012 [hep-th]].


\bibitem{Balasubramanian:1999re}
  V.~Balasubramanian and P.~Kraus,
  {\it A stress tensor for anti-de Sitter gravity},
  {Commun.\ Math.\ Phys.}  {\bf 208} (1999) 413
  [arXiv:hep-th/9902121].

\bibitem{Henningson:1998gx}
  M.~Henningson and K.~Skenderis,
{\it The holographic Weyl anomaly},
  JHEP {\bf 9807} (1998) 023
  [hep-th/9806087];

  M.~Henningson and K.~Skenderis,
  {\it Holography and the Weyl anomaly},
  Fortsch.\ Phys.\  {\bf 48} (2000) 125
  [hep-th/9812032].


\bibitem{Strominger:1997eq}
  A.~Strominger,
  {\it Black hole entropy from near-horizon microstates},
  JHEP {\bf 9802} (1998) 009
  [arXiv:hep-th/9712251].

\bibitem{Carlip:1999cy}
  S.~Carlip,
 {\it Entropy from conformal field theory at Killing horizons},
  Class.\ Quant.\ Grav.\  {\bf 16} (1999) 3327
  [gr-qc/9906126].
 

\bibitem{Brown:1994gs}
  J.~D.~Brown, J.~Creighton and R.~B.~Mann,
{\it Temperature, energy and heat capacity of asymptotically
anti-de Sitter black holes},
  Phys.\ Rev.\ D {\bf 50} (1994) 6394
  [gr-qc/9405007].



\bibitem{Carlip:2005zn}
  S.~Carlip,
{\it Conformal field theory, (2+1)-dimensional gravity, and the BTZ black hole},
  Class.\ Quant.\ Grav.\  {\bf 22} (2005) R85
  [gr-qc/0503022].
 
 

 





 
 
 

 



 
 
 

 



 




\end{thebibliography}
\end{document}